\documentclass[12pt]{iopart}
\usepackage{iopams}  
\usepackage{subfigure,epsfig}
\usepackage{bm}
\usepackage{float}
\usepackage{xspace}
\usepackage{times}
\usepackage{cite}
\usepackage{color}
\newcommand{\bmath}[1]{\ensuremath{\bm{#1}}\xspace}

\newcommand{\x}{\bmath{x}}
\newcommand{\y}{\bmath{y}}

\newcommand{\f}{\bmath{f}}
\newcommand{\g}{\bmath{g}}

\newcommand{\lv}{\bmath{\ell}}

\newcommand{\nv}{\bmath{n}}

\newcommand{\uv}{\bmath{u}}

\newcommand{\pv}{\bmath{p}}
\newcommand{\rv}{\bmath{r}}
\newcommand{\w}{\bmath{w}}

\newcommand{\0}{\bmath{0}}
\newcommand{\1}{\bmath{1}}
\newcommand{\alp}{\bmath{\alpha}}

\newcommand{\lam}{\bmath{\lambda}}
\newcommand{\ome}{\bmath{\omega}}

\newcommand{\ro}{\bmath{\rho}}

\newcommand{\muv}{\bmath{\mu}}
\newcommand{\etav}{\bmath{\eta}}

\newcommand{\phiv}{\bmath{\phi}}

\newcommand{\A}{\bmath{A}}

\newcommand{\G}{\bmath{G}}

\newcommand{\K}{\bmath{K}}

\newcommand{\U}{\bmath{U}}

\newcommand{\beq}{\begin{equation}}
\newcommand{\eeq}{\end{equation}}
\newcommand{\bea}{\begin{eqnarray}}
\newcommand{\eea}{\end{eqnarray}}
\newcommand{\ba}{\left(\!\!\begin{array}}
\newcommand{\ea}{\end{array}\!\!\right)}
\newcommand{\bc}{\begin{center}}
\newcommand{\ec}{\end{center}}

\newcommand{\diag}{\mathrm{diag}}

\newcommand{\true}{\mathrm{true}}

\begin{document}

\title[PET-enabled Dual-Energy CT]{PET-enabled Dual-Energy CT: Image Reconstruction and A Proof-of-Concept Computer Simulation Study}
\author{Guobao Wang}
\address{Department of Radiology, University of California at Davis}
\ead{gbwang@ucdavis.edu}

\begin{abstract}
Standard dual-energy computed tomography (CT) uses two different X-ray energies to obtain energy-dependent tissue attenuation information to allow quantitative material decomposition. The combined use of dual-energy CT and positron emission tomography (PET) may provide a more comprehensive characterization of disease states in cancer and other diseases. However, the integration of dual-energy CT with PET is not trivial, either requiring costly hardware upgrade or increasing radiation dose. This paper proposes a dual-energy CT imaging method that is enabled by the already-available PET data on PET/CT.  Instead of using a second X-ray CT scan with a different energy, this method exploits time-of-flight PET image reconstruction via the maximum likelihood attenuation and activity (MLAA) algorithm to obtain a 511 keV gamma-ray attenuation image from PET emission data. The high-energy gamma-ray CT image is then combined with the low-energy X-ray CT of PET/CT to provide a pair of dual-energy CT images. A major challenge with the standard MLAA reconstruction is the high noise present in the reconstructed 511 keV attenuation map, which would not compromise the PET activity reconstruction too much but may significantly affect the performance of the gamma-ray CT for material decomposition. To overcome the problem, we further propose a kernel MLAA algorithm to exploit the prior information from the available X-ray CT image. We conducted a computer simulation to test the concept and algorithm for the task of material decomposition. The simulation results demonstrate that this PET-enabled dual-energy CT method is promising for quantitative material decomposition. The proposed method can be readily implemented on time-of-flight PET/CT scanners to enable simultaneous PET and dual-energy CT imaging.

\end{abstract}


\section{Introduction}
Dual-energy (DE) computed tomography (CT) has gained increasing popularity in recent years thanks to its capability of differentiating tissue materials \cite{McCollough15}. Different from traditional CT imaging that commonly uses single X-ray energy ($\leq$140 keV), DECT employs two different X-ray energies, one at a lower level (e.g., 50 keV) and the other at a higher level (e.g., 80 keV), to scan the same object either sequentially by two scans or simultaneously by two X-ray sources. It obtains energy-dependent attenuation information of tissue properties and allows quantitative material decomposition \cite{McCollough20}. 

Because it brings a dimension of information that is distinct from what functional positron emission tomography (PET) offers, DECT can complement PET/CT imaging. DECT can be used for improved attenuation correction for PET \cite{Noh09, Xia14} or combined with PET to provide a more comprehensive characterization of diseases \cite{Cecco18, Wu20}. Integration of DECT with PET, however, would not be trivial. Direct replacement of traditional single-energy CT with new DECT is costly because DECT has a different scanner configuration and its price is higher than that of single-energy CT. Utilization of existing single-energy CT scanners is possible but requires significant protocol modification and is also associated with increased radiation exposure and scanning cost \cite{McCollough15}. Another option is sequential two-step PET/DECT imaging on separate scanners, i.e. a PET/CT scan followed by a DECT scan or vice versa. This method, however, has all the disadvantages of separate PET and CT imaging before the invention of integrated PET/CT scanners, including the difficulty of image fusion, extended imaging time, and increased radiation exposure \cite{Townsend08}.

We propose a different dual-energy CT imaging method that is enabled by the already-available PET data on PET/CT instead of using a second X-ray CT scan with a different energy. The method does not require a change of scanner hardware of PET/CT or add additional radiation dose or scan time. It only requires a standard PET/CT scan on a time-of-flight (TOF) PET/CT scanner. The assumption is that a high-energy gamma-ray attenuation image can be reliably obtained from time-of-flight PET emission data. This PET-enabled ``$\gamma$-ray computed tomography (GCT)" image is then combined with the X-ray CT image from PET/CT to produce a pair of dual-energy CT images. 

The theoretical foundation of this idea is supported by the advances in statistical image reconstruction of time-of-flight PET emission scan data for joint estimation of radiotracer activity and attenuation \cite{Defrise12,Rezaei12,Nuyts18}.  Theoretical analysis and practical studies have demonstrated that the gamma-ray attenuation image at 511 keV can be jointly estimated with the reconstruction of PET activity image from time-of-flight PET emission data, for example, using the maximum likelihood activity and attenuation (MLAA) reconstruction algorithm \cite{Nuyts1999, Rezaei12}. Previous attention on this topic has been given to achieve transmission-less PET imaging by excluding the X-ray CT component (e.g., \cite{Defrise14, Rezaei14, Li15, Berker16, Feng18, Cheng20}). Existing studies were also primarily aimed to improve the aspect of attenuation correction for PET activity image reconstruction for PET/CT (e.g., \cite{Panin13, Presotto15, Bousse16, Rezaei18}) or PET/MR (e.g., \cite{Mehranian15, Benoit16, Heuber17, Ahn18, Hwang18, Rezaei19}). The $\gamma$-ray attenuation image itself did not receive much attention and no work explored it for dual-energy or multi-energy CT spectral imaging, which however is the focus of this paper.

One challenge with using the standard MLAA reconstruction to enable the proposed PET-enabled dual-energy CT method is that the estimated GCT image by MLAA is commonly noisy, see \cite{Panin13} for an example. While the noise would not compromise the performance significantly if the usage is for PET attenuation correction, it may largely affect the quantitative accuracy of the GCT for multi-material decomposition. 

To suppress the noise, we note that the GCT image shares the same anatomical structures as the X-ray CT image because both reflect the linear attenuation maps, though at two different energies. Therefore, we propose to utilize the already available X-ray CT image as the {\em a priori} information to guide the reconstruction of GCT from the PET data. To incorporate image prior, previous image reconstruction methods commonly employ an explicit regularization form (e.g., \cite{Bowsher96, Mehranian15}) which can be complex for practical implementation. Regularization-based methods also often require a convergent solution to achieve the optimal performance, which is computationally costly. In comparison, the kernel method \cite{Wang2015, Hutchcroft16, Novosad16, Bland18, Gong18, Deidda19, Wang2019} encodes image prior information in the forward model of tomographic image reconstruction and requires no explicit regularization. It is easier to implement and can be more efficient and better improve PET image reconstruction than regularization-based methods \cite{Wang2015,Hutchcroft16}. In this work, we adopt the kernel framework and develop a kernel MLAA algorithm to incorporate the X-ray CT image prior knowledge for noise suppression in the MLAA attenuation image reconstruction. 

Part of this work was presented in the 2018 IEEE Nuclear Science Symposium and Medical Imaging Conference \cite{Wang18}. Compared to its conference version, this paper has been substantially extended by including the development and validation of the kernel MLAA algorithm to solve the noise challenge and a more comprehensive computer simulation study to demonstrate the feasibility of the proposed PET-enabled dual-energy CT method. 


\begin{figure}[t]
\bc
{\includegraphics[trim=0cm 0cm 0cm 0cm, clip, width=10cm]{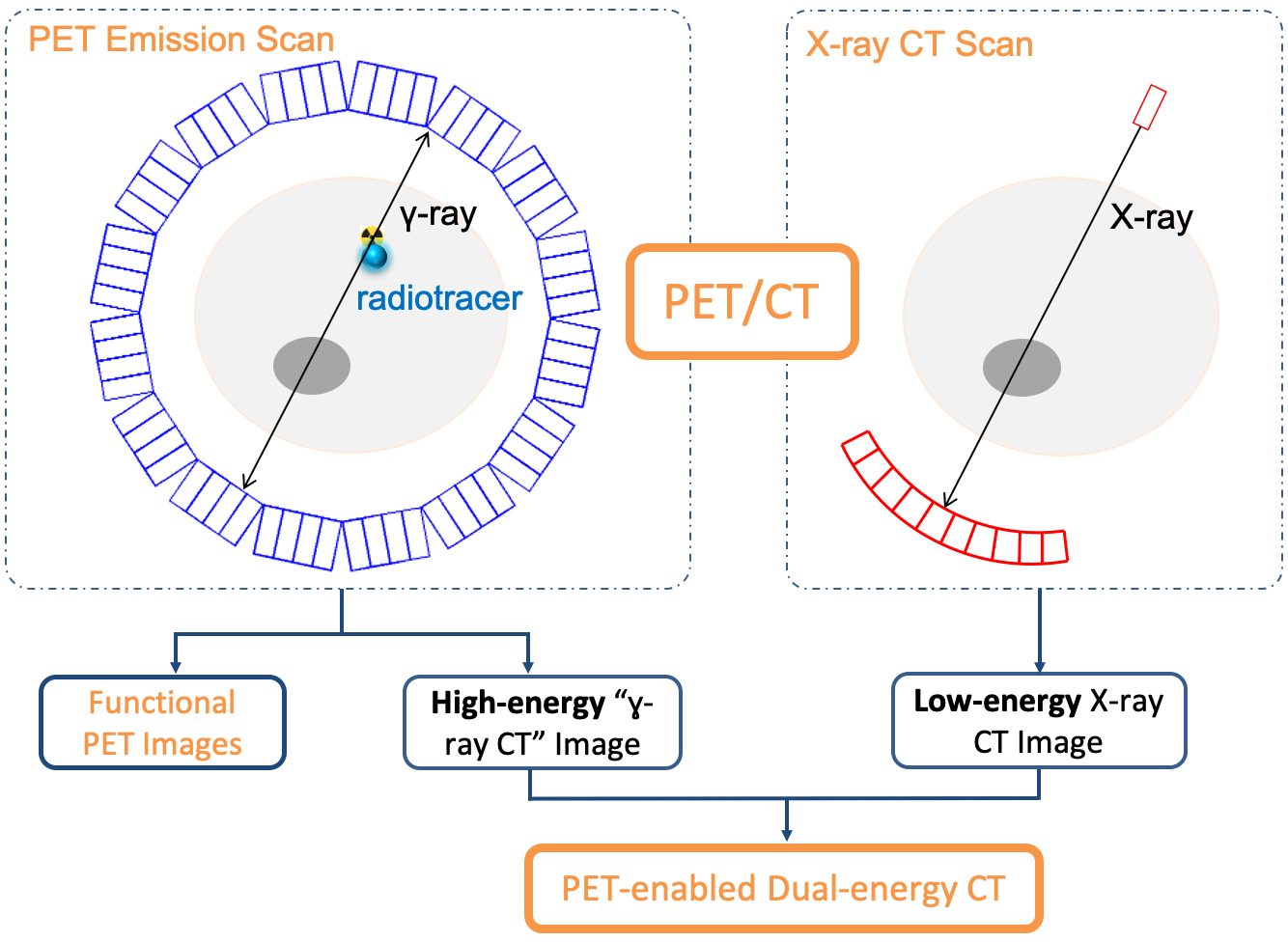}}
\caption{The PET-enabled dual-energy CT imaging method exploits the annihilation radiation of PET radiotracer decays as an internal ``$\gamma$-ray" source to reconstruct a high-energy attenuation image from PET emission data, which is then combined with the low-energy X-ray CT image to form a dual-energy CT image pair.}
\label{fig-mmd-img}
\ec
\end{figure}

\section{Proposed PET-enabled Dual-energy CT Method}

\subsection{The Idea}

As illustrated in Fig. 1, a standard PET/CT scan normally consists of a PET emission scan at 511 keV and a X-ray CT transmission scan commonly acquired at 80-140 kVp. The X-ray CT image has been mainly used for PET attenuation correction with which the PET scan provides a functional image describing the radiotracer distribution in the subject. Our proposed method exploits the potential of a standard PET emission scan for high-energy GCT imaging. Different from X-ray CT which uses an external X-ray source to generate tomographic data, here the PET-enabled GCT exploits the internal ``$\gamma$-rays" generated by annihilation radiation of PET radiotracer decay in the object. The GCT image obtained from PET is then combined with the low-energy X-ray CT to form dual-energy CT imaging. 

There are potentially multiple methods for obtaining a GCT image. In this work, we exploit a PET attenuation-activity joint reconstruction method, as described below.

\subsection{Joint Reconstruction of GCT Image from PET Data}

PET projection measurement $\y$ can be well modeled as independent Poisson random variables with the log-likelihood function,
\beq
L(\y|\lam,\muv)=\sum_{i=1}^{N_d}\sum_{m=1}^{N_t} y_{i,m} \log \bar{y}_{i,m}(\lam,\muv)-\bar{y}_{i,m}(\lam,\muv),
\eeq
where $i$ denotes the index of PET detector pair and $m$ denotes the index of time-of-flight (TOF) bin. $N_d$ is the total number of detector pairs and $N_t$ is the number of TOF bins. The expectation of the PET projection data $\bar{\y}(\lam,\muv)$ is related to the activity image $\lam$ and  attenuation image $\muv$ at 511 keV through 
\beq
\bar{\y}_{m}(\lam,\muv)= \diag\left\{\nv_m(\muv)\right\}\G_m\lam + \rv_{m}
\label{eq-forwproj}
\eeq
where $\G_m$ is the PET detection probability matrix for the $m$th timing bin and $\rv$ accounts for the expectation of random and scattered events. $\nv_m(\muv)$ is the normalization factor for TOF bin $m$, of which the $i$th element is 
\beq
n_{i,m}(\muv)=c_{i,m}\cdot\exp{\big(-[\A\muv)]_i\big)}
\eeq
where $c_{i,m}$ denotes the multiplicative factor other than the attenuation correction factor and $\A$ is the system matrix for transmission imaging. 

For standard PET/CT imaging, the attenuation image $\muv$ is normally predetermined from a X-ray CT scan and the PET reconstruction problem only estimates the $\lam$ image \cite{Qi2006}. $\muv$ can be approximated from a X-ray CT scan using a bilinear scaling conversion of linear attenuation coefficient from the X-ray energy (e.g., 140 kVp) to 511 keV \cite{Kinahan03}.

The maximum-likelihood attenuation and activity (MLAA) estimation method \cite{Nuyts1999, Rezaei12} seeks the estimates of both $\muv$ and $\lam$ simultaneously by maximizing the Poisson log-likelihood,
\beq
\hat{\lam},\hat{\muv}=\arg\max_{\lam\geq\0,\muv\geq\0} L(\y|\lam,\muv).
\eeq 
The MLAA formulation was first proposed for non-TOF data \cite{Nuyts1999} but the simultaneous estimation suffers from cross-talk artifacts despite some encouraging results \cite{Nuyts1999, Dicken1999}. The method was later demonstrated more effective for TOF data \cite{Salomon2011, Conti2011}. 
A seminal theoretical work later proved that TOF data determine $\muv$ up to a constant \cite{Defrise12, Rezaei12}. Since then, the MLAA method has received a wide range of interests
(e.g., \cite{Panin13, Defrise14, Rezaei14, Mehranian15, Li15, Presotto15, Bousse16, Berker16, Benoit16, Heuber17, Feng18, Rezaei18, Ahn18, Hwang18, Rezaei19, Cheng20}). 

It is worth noting that previous attention of MLAA reconstruction was focused on PET attenuation correction.
In this paper, we exploit the MLAA reconstruction distinctly. We propose to combine the GCT image $\muv$ with the X-ray CT image to obtain a dual-energy CT image pair to enable multi-material decomposition.

\subsection{Multi-material Decomposition (MMD)}

For each image pixel $j$, the high-energy GCT attenuation value $\mu_j$ and the low-energy X-ray CT attenuation value $x_j$ form a pair of dual-energy measurements $\uv_j$. The tissue compositions are then described by a set of material bases, for example, air (A), soft tissue (S) or equivalently water, and bone (B):

\beq
\uv_j\triangleq
\ba{c}
x_j\\
\mu_j
\ea
=\underbrace{\ba{c c c c}
x_A &x_S & x_B\\
\mu_A &\mu_S & \mu_B
\ea}_{\U}
\underbrace{\ba{c}
\rho_{j,A}\\ \rho_{j,S} \\ \rho_{j,B}
\ea}_{\ro_j}
\eeq
where the coefficients $\rho_{j,k}$ with $k=\{A,S,B\}$ are the fraction of each basis material in pixel $j$ and subject to 
\beq
\sum_k \rho_{j,k} = 1.
\eeq
The material basis matrix $\U$ consists of the linear attenuation coefficients of each basis material measured at the low and high energies. The estimates of $\ro_j$ are obtained using the following least-square optimization for each image pixel,
\beq
\hat{\ro}_j= \arg\min_{\ro_j} ||\uv_j-\U\ro_j||^2.
\eeq

\section{Improved GCT Reconstruction Using Kernel MLAA}

\subsection{Use of X-ray CT Image Prior}

\begin{figure}[t]
\bc
{\includegraphics[trim=0.2cm 0.1cm 0.1cm 0.1cm, clip, width=8cm]{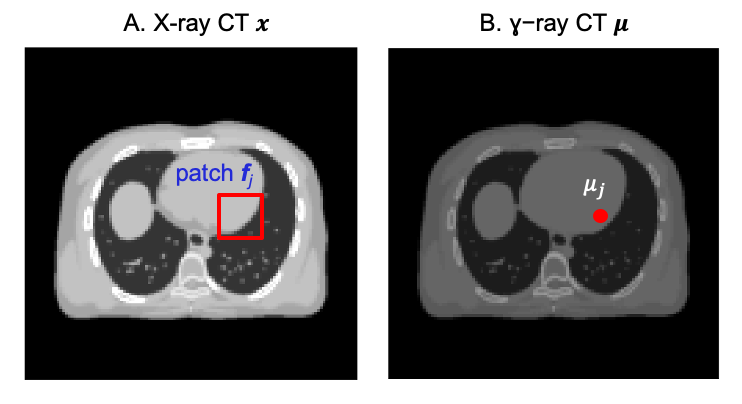}}
\caption{Patches $\{\f_j\}$ extracted from a X-ray CT image $\x$ can be used to build the kernel representation of the intensity in pixel $j$ of the $\gamma$-ray CT image $\muv$. }
\label{fig-xkernel}
\ec
\end{figure}

The GCT by standard MLAA reconstruction is commonly noisy. To suppress the noise, we propose to utilize available X-ray CT in PET/CT as an image prior. As illustrated in figure \ref{fig-xkernel}, 
the higher contrast and potentially much better image quality provided by a X-ray CT image can be beneficial to guide the reconstruction of a $\gamma$-ray CT image.  In this work, we apply the kernel method which was originally developed for dynamic PET reconstruction (e.g., \cite{Wang2015}) and dual-modality imaging such as PET/MR (e.g., \cite{Hutchcroft16,Bland18}). Here we extend the kernel method to exploit low-energy X-ray CT image prior for reconstruction of 511 keV GCT image from PET emission data. 

\subsection{Kernel Representation of  GCT Image}

With the X-ray CT image $\x$, we can extract a ``data point" $\f_j$ for each pixel $j$ from the image, for example, using the image patch centered at $j$ (Fig. \ref{fig-xkernel}). A transformed feature space can be defined by a nonlinear mapping function $\phiv$, which transforms the low-dimensional space $\{\f_j\}$ to a very-high dimensional space $\{\phiv(\f_j)\}$. In the high-dimensional feature space, the intensity of the GCT in pixel $j$ can be described as a linear function,
$
\mu_j = \w^T\phiv(\f_j),
$
where $\w$ denotes the coefficient vector. Because $\w$ also sits in the feature space, i.e., $\w=\sum_l \alpha_l\phiv(\f_l)$, we then have the following equivalent kernel representation for $\mu_j$,
\beq
\mu_j = \sum_l \alpha_l \underbrace{\phiv(\f_j)^T\phiv(\f_l)}_{\kappa(\f_j, \f_l)}
\eeq
where the kernel function $\kappa(\cdot,\cdot)$ is defined as the inner product of the two feature vectors $\phiv(\f_j)$ and $\phiv(\f_l)$. The form of the kernel function can be directly defined without knowing the specific form of $\phiv$.  For example, the radial Gaussian kernel is
\beq
\kappa(\f_j, \f_k) = \exp\left(-||\f_j-\f_k||^2/2\sigma^2\right)
\eeq
which corresponds to a $\phiv$ of infinite dimension. $\sigma$ is a hyper-parameter. 

The matrix-vector form of the kernel representation for the GCT image is 
\beq
\muv=\K\alp.
\label{eq-muker}
\eeq
where $\K$ is the kernel matrix built on the X-ray image $\x$ with its $(j,l)$th element equal to $\kappa(\f_j, \f_l)$. The unknown parameter vector $\alp$ denotes the corresponding kernel coefficient image. Although with a large matrix size, $\K$ can be built to be sparse to make a practical implementation \cite{Wang2015}.

\subsection{Kernel MLAA}

Inserting the kernel representation in  Eq. (\ref{eq-muker}) into the original MLAA formulation leads to a kernelized optimization problem as follows, 
\beq
\hat{\lam},\hat{\alp}=\arg\max_{\lam\geq\0,\alp\geq\0} L\big(\y|\lam,\K\alp\big).
\eeq 
Once $\hat{\alp}$ is obtained, the final estimate of the GCT image is obtained by 
\beq
\hat{\muv}=\K\hat{\alp}.
\eeq

To solve the optimization problem, we use the same alternating optimization strategy as used in \cite{Rezaei12}. Each iteration of the algorithm consists of two separate $\lam$-step and $\alp$-step,

\bea
\hat{\lam}&=&\arg\max_{\lam\geq\0} L\big(\y|\lam,\K\hat{\alp}\big),\\
\hat{\alp}&=&\arg\max_{\alp\geq\0} L\big(\y|\hat{\lam},\K\alp\big).
\eea

\subsubsection{$\lam$-estimation step}

The $\lam$-step is a maximum-likelihood PET reconstruction problem which can be easily solved using the standard expectation-maximum (EM) algorithm,
\beq
\lam^{n+1}=\frac{\lam^{n}}{\hat{\pv}}\cdot\left(\sum_m \G_m^T \left[\nv_m(\hat{\muv})\cdot\frac{\y_m}{\bar{\y}_m(\lam^n,\hat{\muv})}\right]\right)
\eeq
where $n$ denotes the inner iteration number and the superscript $T$ denotes matrix or vector transpose. $\hat{\pv}$ is the sensitivity image defined by
\beq
\hat{\pv}=\sum_m \G_m^T\nv_m(\hat{\muv}).
\eeq

\subsubsection{$\alp$-estimation step}

The $\alp$-step is a kernel maximum-likelihood transmission reconstruction (MLTR) problem \cite{Erdogan99} for time-of-flight PET data,
\beq
\hat{\alp}=\arg\max_{\alp\geq\0} \sum_{i=1}^{N_d}\sum_{m=1}^{N_t} \left[-h_{i,m}\Big([\A\K\alp]_i\Big)\right],
\eeq
where $h_{i,m}(\ell)$ is the negative likelihood function,
\beq
h_{i,m}(\ell)\triangleq  \big(\hat{b}_{i,m}e^{-\ell}+r_{i,m}\big) - y_{i,m}\log\big(\hat{b}_{i,m}e^{-\ell}+r_{i,m}\big),
\eeq
with $\hat{b}_{i,m}=c_{i,m} \cdot[\G_m\hat{\lam}]_i$.

The kernel MLTR problem can be solved using the optimization transfer principle in  \cite{Erdogan99} to construct the following quadratic surrogate, 
\beq
Q(\alp;\alp^n)=-||\hat{\lv}^{n+1} - \A\K\alp||_{\hat{\etav}^{n}}^2,
\eeq
where $\hat{\lv}^{n+1}$ is an intermediate GCT sinogram,
\beq
\hat{\ell}^{n+1}_i =\ell_i^n-\frac{\sum_m\dot{h}_{i,m}(\ell_i^n)}{\sum_m \eta_{i,m}(\ell_i^n) },
\eeq 
and $\hat{\etav}^n$ is an intermediate weight sinogram,
\beq
\hat{\eta}_i^n=\sum_m \eta_{i,m}(\ell_i^n).
\eeq

In the equations, $\ell_i^n=[\A\muv^n]_i$ with $\muv^n = \K\alp^n$. $\eta_{i,m}(\ell_i^n)$ is the optimum curvature defining a quadratic surrogate function that majorizes the function $h_{i,m}(\ell)$ \cite{Erdogan99}, 
\beq
\eta_{i,m}(\ell) = \left\{
\begin{array}{ll}
\frac{2}{\ell^2}\left[h_{i,m}(0)-h_{i,m}(\ell)+\ell\dot{h}_{i,m}(\ell)\right]_+,&\ell > 0,\\
\left[\ddot{h}_{i,m}(\ell)\right]_+,&\ell = 0,
\end{array} \right.
\eeq
where $[\cdot]_+=\max(0,\cdot)$ applies the non-negativity constraint. $\dot{h}$ and $\ddot{h}$ are the the first and second derivatives of $h_{i,m}(\ell)$, respectively \cite{Erdogan99}.

The surrogate function $Q(\alp;\alp^n)$ minorizes the original likelihood function $L$ and meets,
\beq
Q(\alp;\alp^n)-Q(\alp^n;\alp^n) \leq L\big(\y|\hat{\lam},\K\alp\big) - L\big(\y|\hat{\lam},\K\alp^n\big)
\eeq
\beq
\nabla Q(\alp;\alp^n) = \nabla L\big(\y|\hat{\lam},\K\alp\big). 
\eeq
where $\nabla$ denotes the gradient with respect to $\alp$.

By treating ``$\A\K$'' as a single matrix, maximization of $Q(\alp;\alp^n)$ can be solved using the separable quadratic surrogate (SQS) algorithm  \cite{Erdogan99b},
\beq
\alp^{n+1}=\left[\alp^{n} - \frac{\g^{n}}{\ome^{n}} \right]_+
\label{eq-sps}
\eeq
where $\g^n$ is the gradient of $Q(\alp;\alp^n)$,
\beq
\g^n=\K^T\A^T\diag(\hat{\etav}^n)\A\K\big(\hat{\lv}^{n+1}-\lv^n\big)
\eeq
and $\ome^n$ is an intermediate weight image,
\beq
\ome^n = \K^T\A^T\diag(\hat{\etav}^n)\A\K\1
\eeq
with $\1$ denoting the all-one vector.

Following the optimization transfer principle \cite{Erdogan99}, the update given by Eq. (\ref{eq-sps}) is guaranteed to monotonically increase the Poisson log likelihood, i.e.,
\beq
L\big(\y|\hat{\lam},\K\alp^{n+1}\big)\geq L\big(\y|\hat{\lam},\K\alp^n\big).
\eeq

\section{Validation Using Computer Simulation}

\subsection{Computer Simulation Setup}

\begin{figure}[t]
\bc\footnotesize
\subfigure[]{\includegraphics[trim=2cm 1cm 1cm 0cm, clip, width=4.9cm]{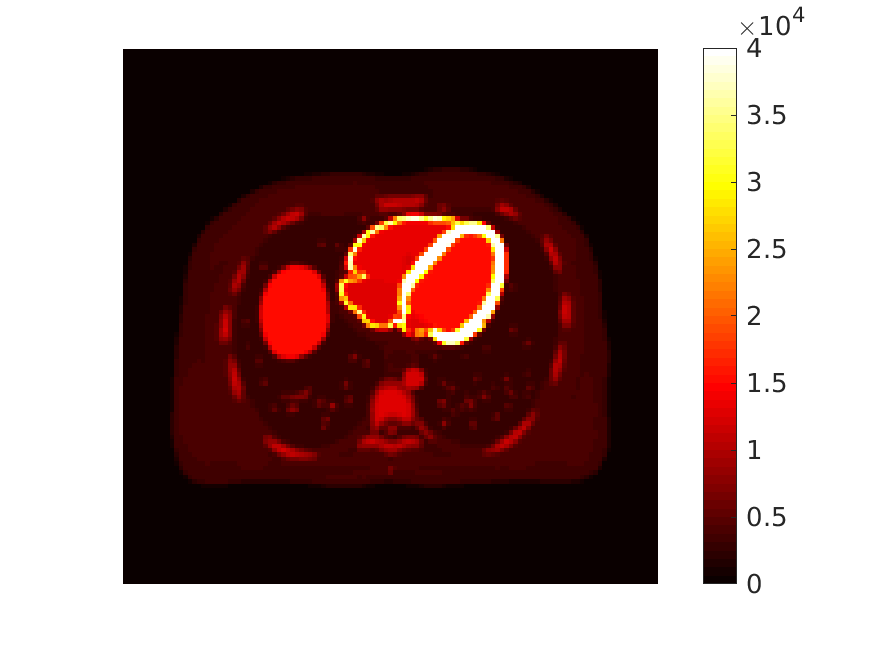}}
\subfigure[]{\includegraphics[trim=2cm 1cm 1cm 0cm, clip, width=5.0cm]{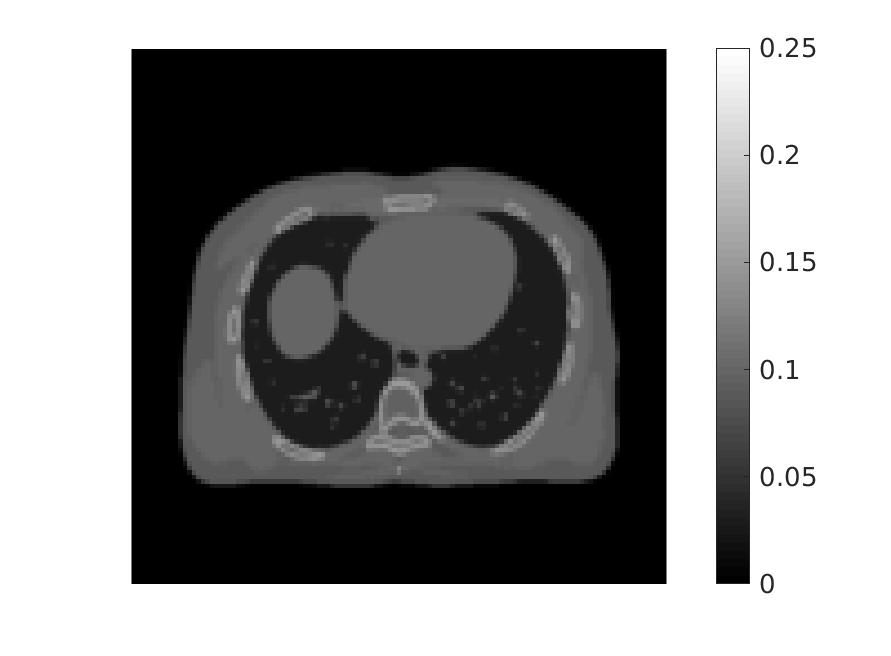}}
\label{fig-sim-images}
\subfigure[]{\includegraphics[trim=2cm 1cm 1cm 0cm, clip, width=5.0cm]{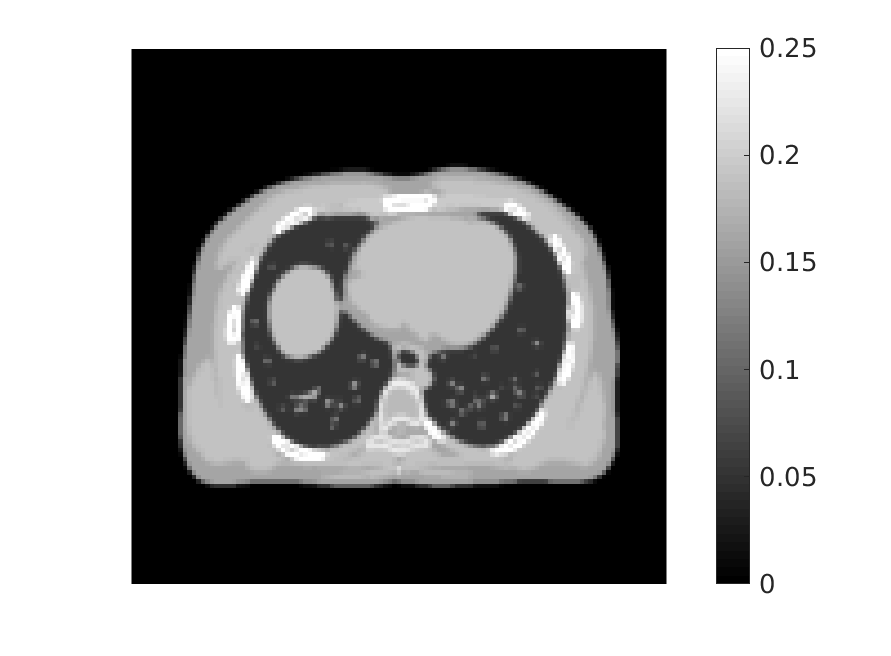}}
\caption{The digital phantom used in the PET/CT computer simulation. (a) PET activity image in Bq/cc; (b) PET attenuation image at 511 keV in cm$^{-1}$; (c) X-ray CT image at 80 keV. }
\label{fig-pht-xcat}
\ec
\end{figure}

We simulated the GE Discovery 690 PET/CT scanner in 2D. The TOF timing resolution of this PET scanner was about 550 ps. The simulation was conducted using the XCAT phantom. The true PET activity image and 511 keV attenuation image are shown in Fig. \ref{fig-pht-xcat}(a) and (b), respectively. The images were first forward projected to generate noise-free PET sinogram of 11 TOF bins. A 40\% uniform background was included to simulate random and scattered events. Poisson noise was then generated using 5 million expected events, unless specified otherwise. The x-ray CT image at a low-energy 80 keV is shown in Fig. \ref{fig-pht-xcat}(c). The data were reconstructed into images of $180\times180$ with a pixel size of $3.9\times 3.9$ mm$^2$.

\subsection{Reconstruction Methods to Compare}

Three reconstruction algorithms were compared in this study: (1) the standard MLAA algorithm, (2) proposed kernel MLAA, and (3) post-reconstruction kernel smoothing using the same kernel matrix $\K$. The third algorithm is also equivalent to nonlocal means denoising \cite{Wang2015}.  Using the $3\times 3$ image patches extracted from the X-ray CT image $\x$, the kernel matrix was built using 50 nearest neighbors in a way similar to \cite{Wang2015}. Each of the standard MLAA and kernel MLAA algorithms was run for 3000 iterations. Within each iteration, one inner iteration was used for the PET activity $\lam$ estimation step and five inner iterations were used for the attenuation $\muv$ estimation step.

For each reconstruction algorithm, two different initial image estimates were used for the GCT reconstruction. One is the uniform initial with $\mu_j^{\mathrm{init}}=0.1~\textrm{cm}^{-1}$ and the other is the 511 keV attenuation map converted from the X-ray CT image using a bilinear scaling.

\begin{figure*}[t]
\bc\footnotesize
\begin{tabular}{c c c}
&{\includegraphics[trim=2cm 1cm 1.3cm 0cm, clip, height=4.1cm]{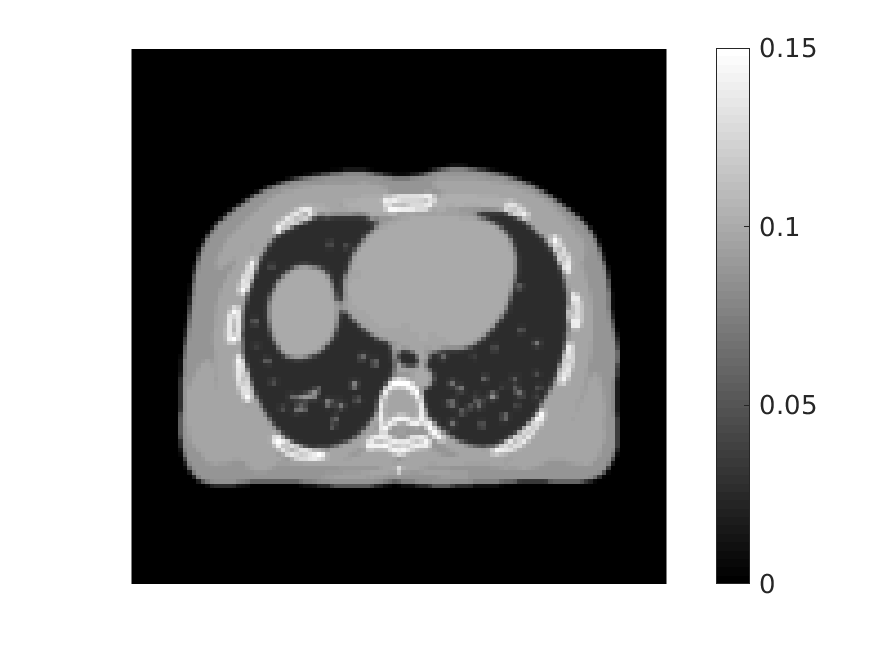}} &\\
&(a)&\\
{\includegraphics[trim=2cm 1cm 3.5cm 0cm, clip, height=4.1cm]{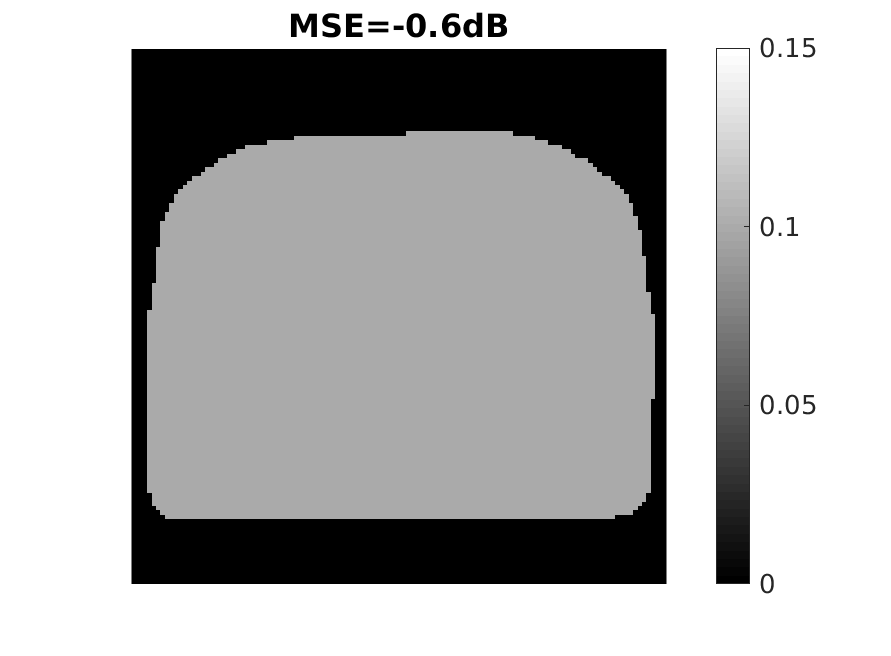}}&
{\includegraphics[trim=2cm 1cm 3.5cm 0cm, clip, height=4.1cm]{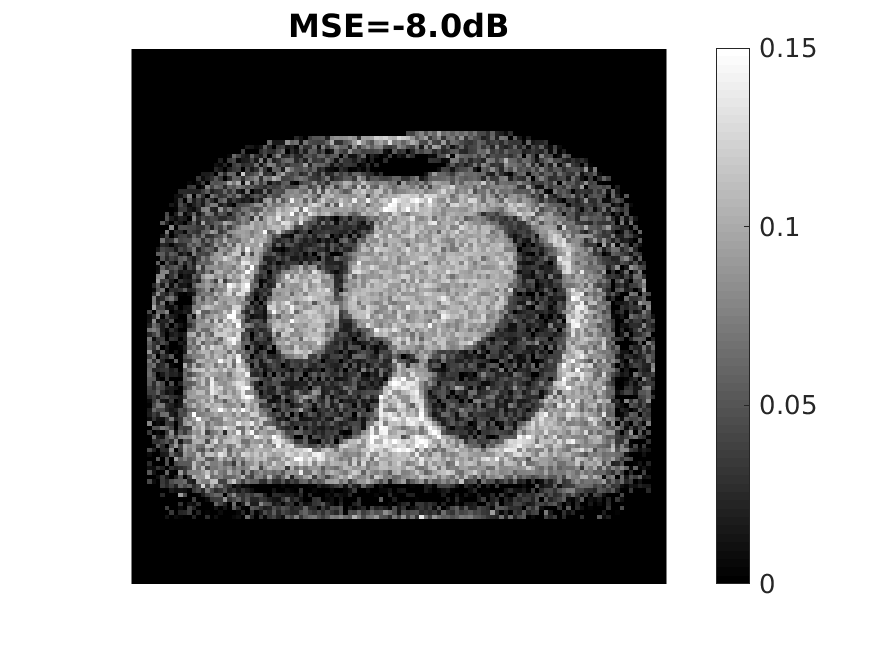}}&
{\includegraphics[trim=2cm 1cm 1.3cm 0cm, clip, height=4.1cm]{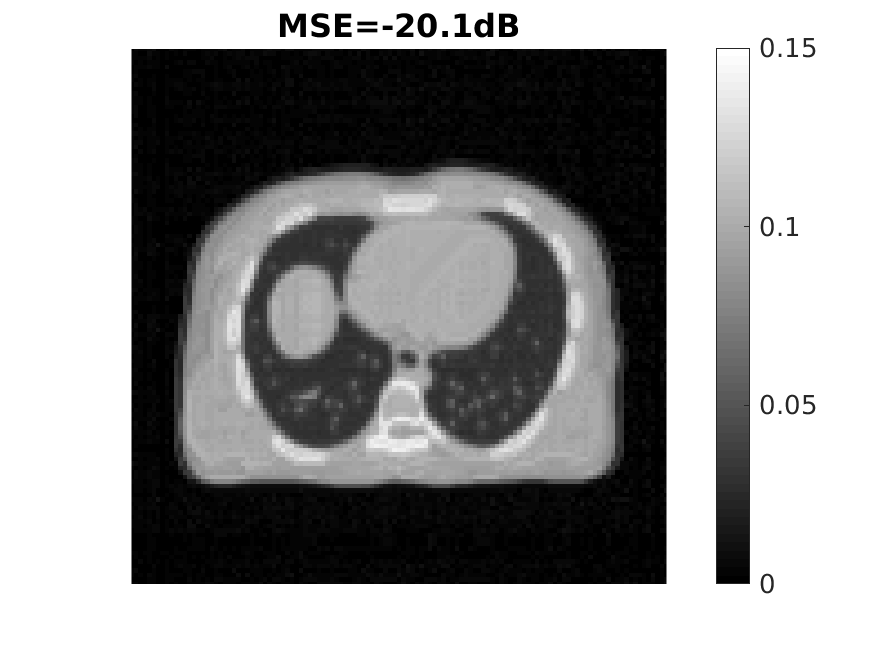}}\\
(b)& (c) & (d)\\
{\includegraphics[trim=2cm 1cm 3.5cm 0cm, clip, height=4.1cm]{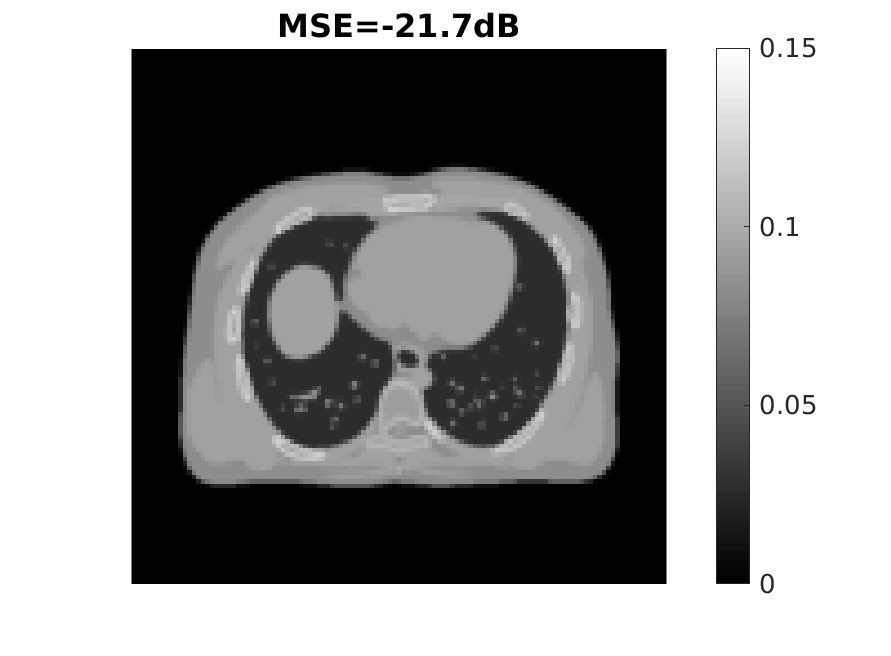}}&
{\includegraphics[trim=2cm 1cm 3.5cm 0cm, clip, height=4.1cm]{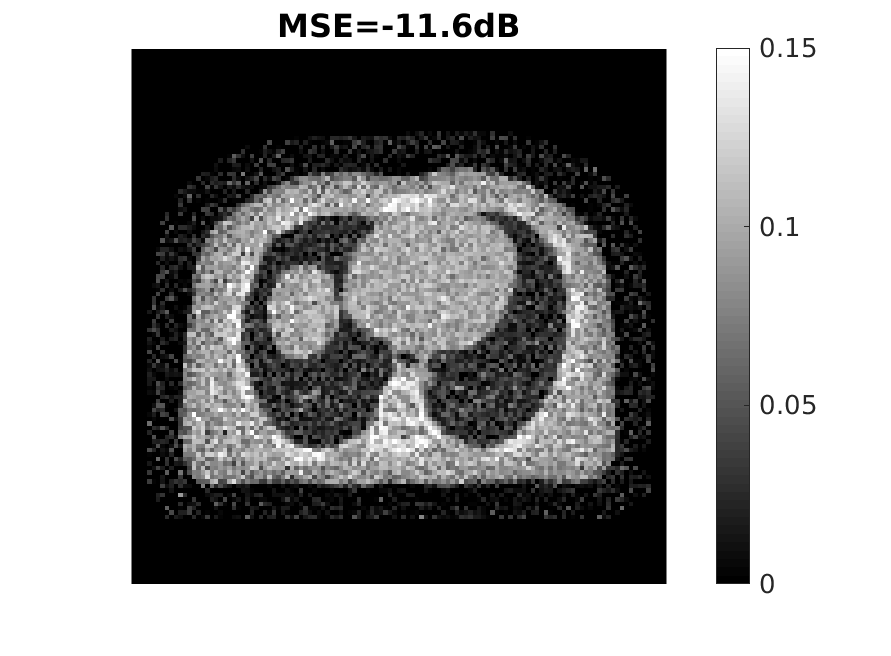}}&
{\includegraphics[trim=2cm 1cm 1.3cm 0cm, clip, height=4.1cm]{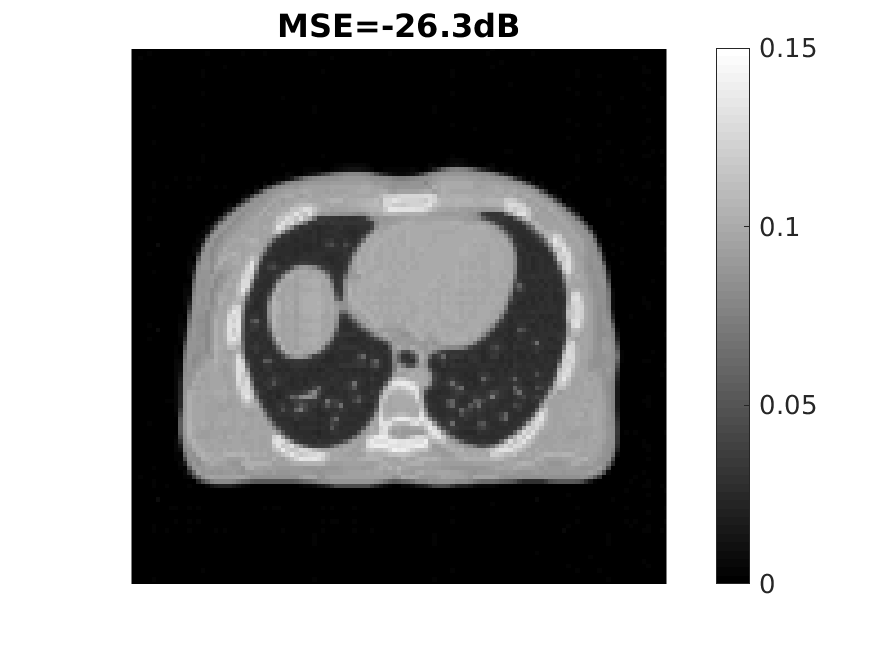}}\\
 (e)& (f) & (g)
\end{tabular}
\caption{Reconstructed GCT images by different reconstruction algorithms and initial estimates. (a) ground truth; (b-d): uniform initial (b) and the standard MLAA (c) and proposed kernel MLAA (d) reconstructions; (e-g): the X-ray CT converted initial estimate at 511 keV (e) and the corresponding standard MLAA (f) and proposed kernel MLAA (g) estimates. 400 iterations were used for the reconstructions.}
\label{fig-rec-images}
\ec
\end{figure*}

\begin{figure*}[t]
\bc\footnotesize
\subfigure[]{\includegraphics[trim=0.5cm 0cm 1cm 0.5cm, clip, height=7.0cm]{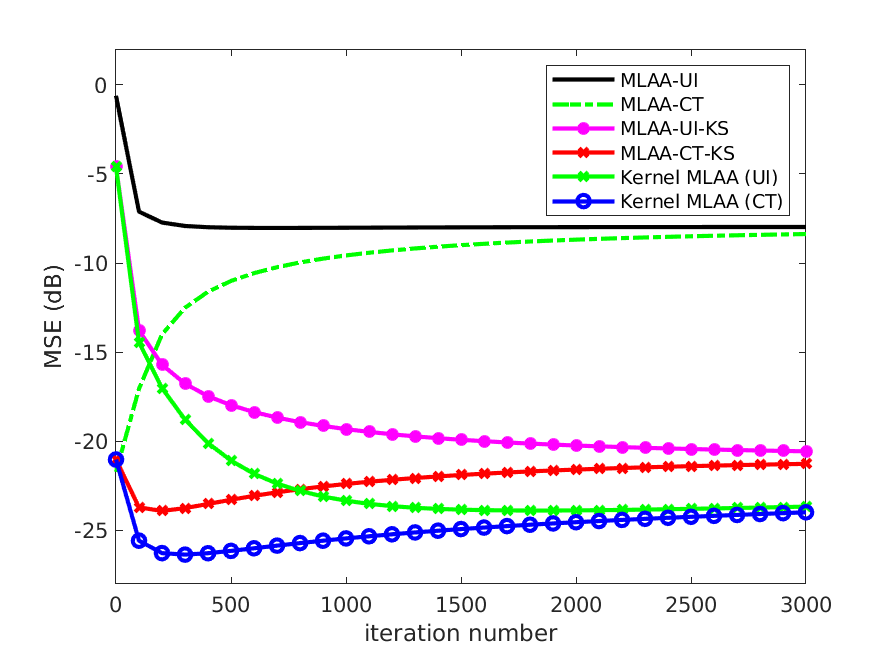}}
\caption{Plot of image MSE as a function of iteration for three different reconstruction algorithms [standard MLAA reconstruction, post-reconstruction denoising with kernel smoothing (KS), and kernel MLAA reconstruction] with two different image initials [uniform initial (UI) and X-ray CT-converted 511 keV attenuation map].}
\label{fig-rec-mse}
\ec
\end{figure*}

\subsection{Evaluation Metrics}

Our main interest is in dual-energy CT imaging. Hence we did not specifically evaluate the algorithms for PET activity reconstruction but focused on the evaluation for the CT performance. 

The quality of GCT was first assessed using the image mean squared error (MSE) defined by
\beq
\mathrm{MSE}(\hat{\muv})=10\log_{10}\frac{||\hat{\muv}-\muv^\true||^2}{||\muv^\true||^2}\quad (\mathrm{dB}),
\label{eq-mse}
\eeq
where $\hat{\muv}$ is an image estimate of GCT obtained with one of the MLAA reconstruction methods and $\muv^\true$ denotes the ground truth GCT image. 

For evaluating quantification, we also calculated the ensemble bias and standard deviation (SD) of the mean intensity in regions of interest (ROI) by
\bea
\mathrm{Bias} &=& \frac{1}{c^\true}|\bar{c}-c^\true|,\\
\mathrm{SD} &=& \frac{1}{c^\true}\sqrt{\frac{1}{N_r-1}\sum_{i=1}^{N_r}|c_i-\bar{c}|^2},
\eea
where $c^\true$ is the noise-free regional intensity and $\bar{c}=\frac{1}{N_r}\sum_{i=1}^{N_r}c_i$ denotes the mean of $N_r$ realizations. $N_r=5$ in this study. 

In addition to the comparison for GCT image quality, different reconstruction algorithms were further compared for dual-energy CT multi-material decomposition as formulated in Section 2. Image MSE, ROI bias and SD were calculated for each of the material basis fraction images. 

\begin{figure}[t]
\bc\footnotesize
{\includegraphics[trim=0cm 0cm 1cm 0cm, clip, height=6.5cm]{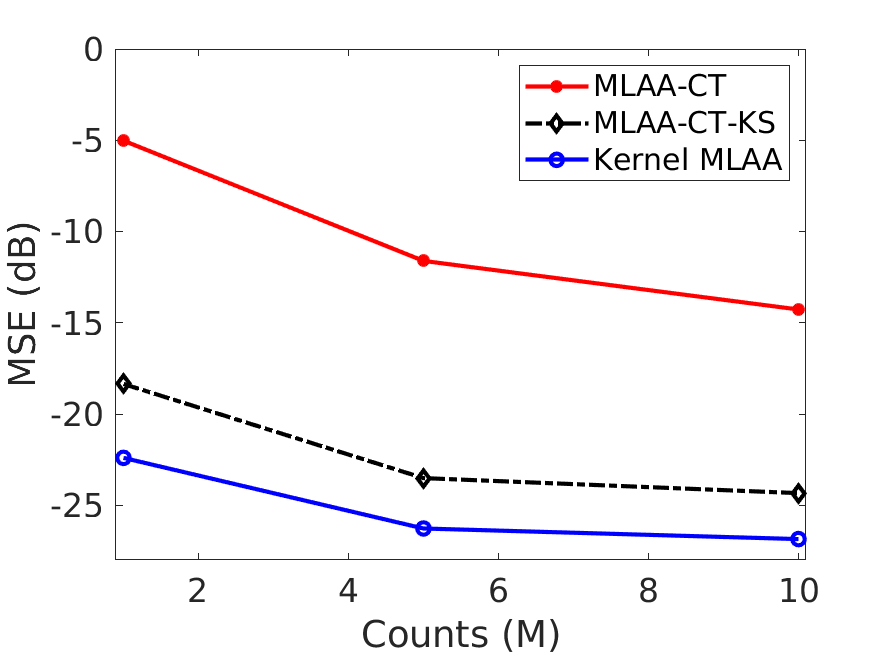}}
\caption{Effect of count level on image MSE for different reconstruction algorithms.}
\label{fig-rec-mse-count}
\ec
\end{figure}

\begin{figure*}[htp]
\bc\footnotesize
\subfigure[]{\includegraphics[trim=2.5cm 2.0cm 1.9cm 1cm, clip, width=5.5cm]{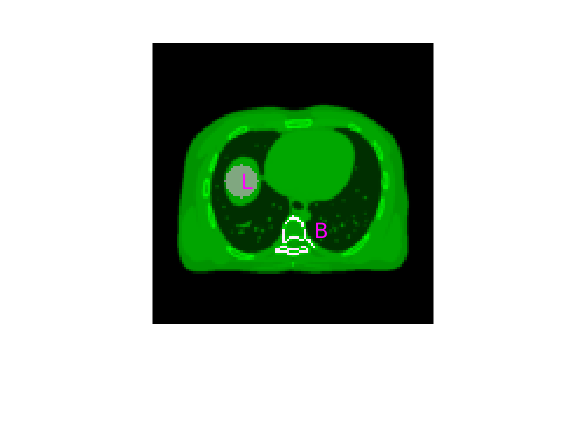}}\\
\subfigure[]{\includegraphics[trim=0cm 0cm 1cm 0cm, clip, height=5.5cm]{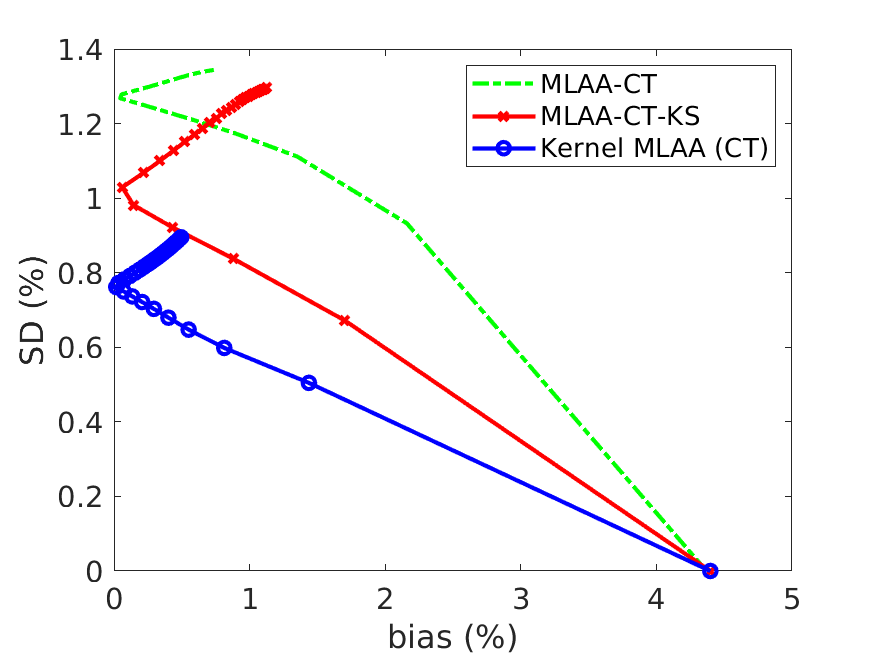}}
\subfigure[]{\includegraphics[trim=0cm 0cm 1cm 0cm, clip, height=5.5cm]{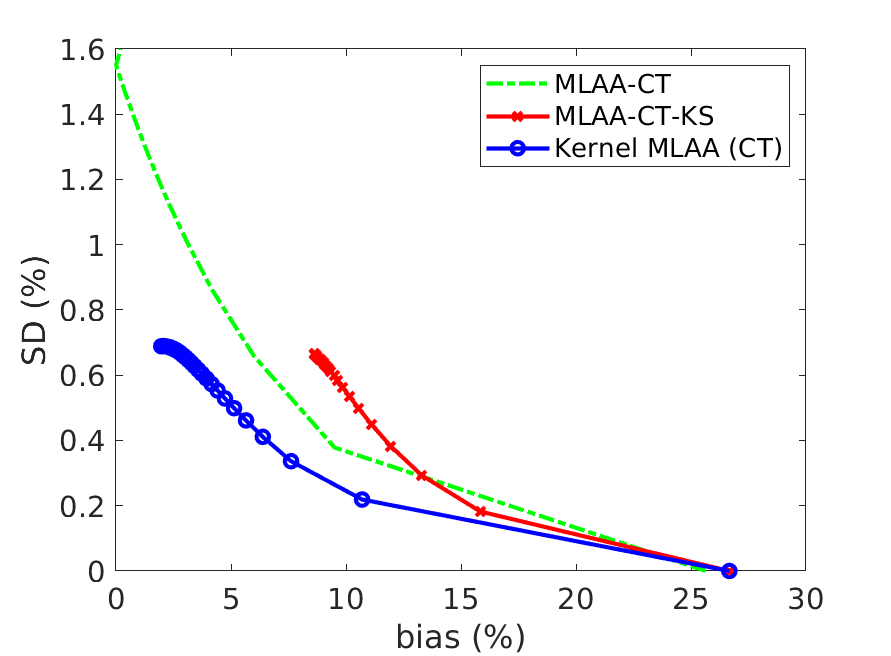}}
\caption{Plot of bias versus standard deviation trade-off for GCT ROI quantification. (a) Illustration of a liver ROI ``L" and a spine bone ROI ``B"; (b) Result of the liver quantification; (c) Result of the bone quantification.}
\label{fig-rec-BiasSD}
\ec
\end{figure*}

\begin{figure*}[t]
\bc\footnotesize
\begin{tabular}{c c c}
{\includegraphics[trim=2cm 1cm 3.5cm 0cm, clip, height=4.2cm]{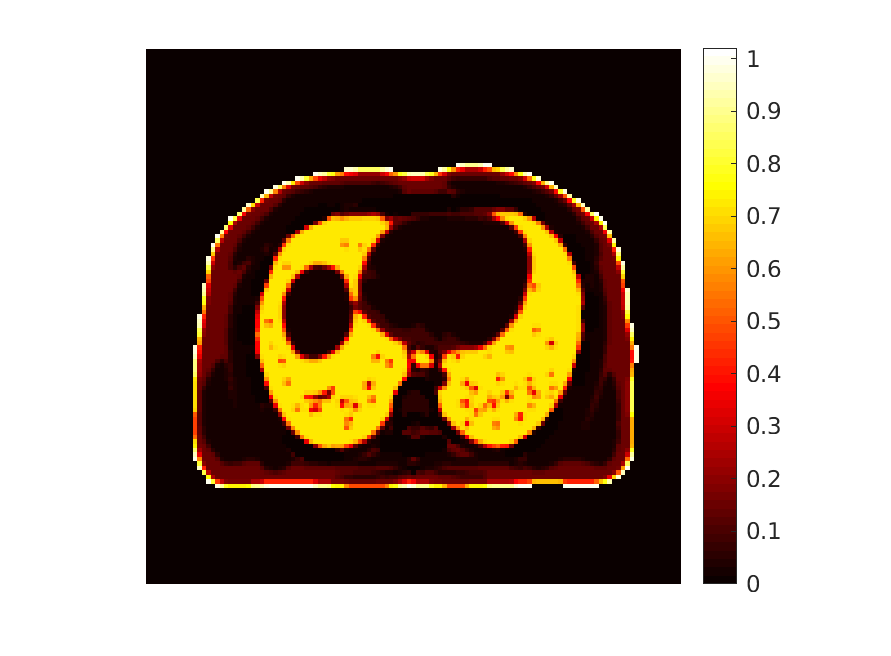}} &
{\includegraphics[trim=2cm 1cm 3.5cm 0cm, clip, height=4.2cm]{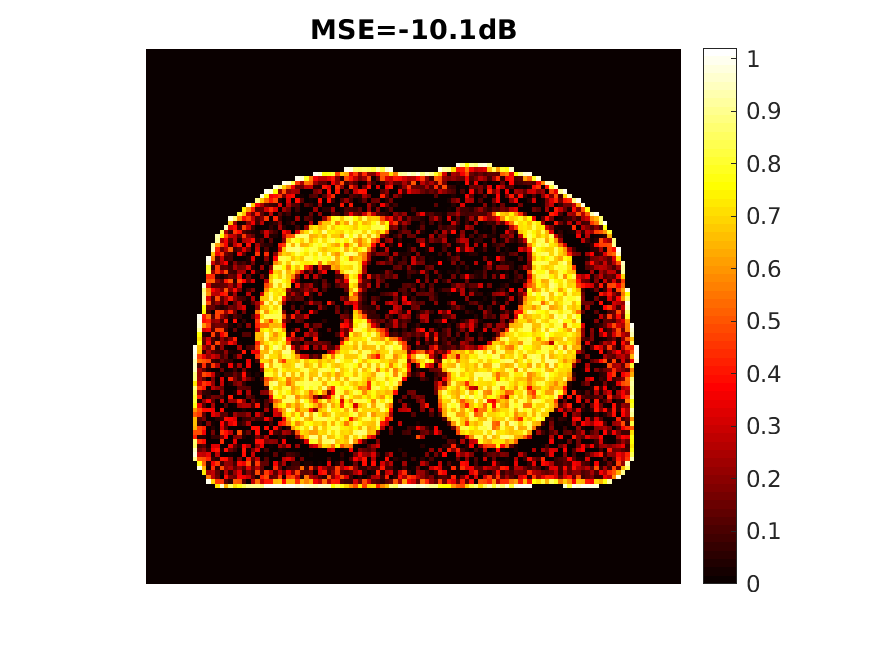}} &
{\includegraphics[trim=2cm 1cm 1.8cm 0cm, clip, height=4.2cm]{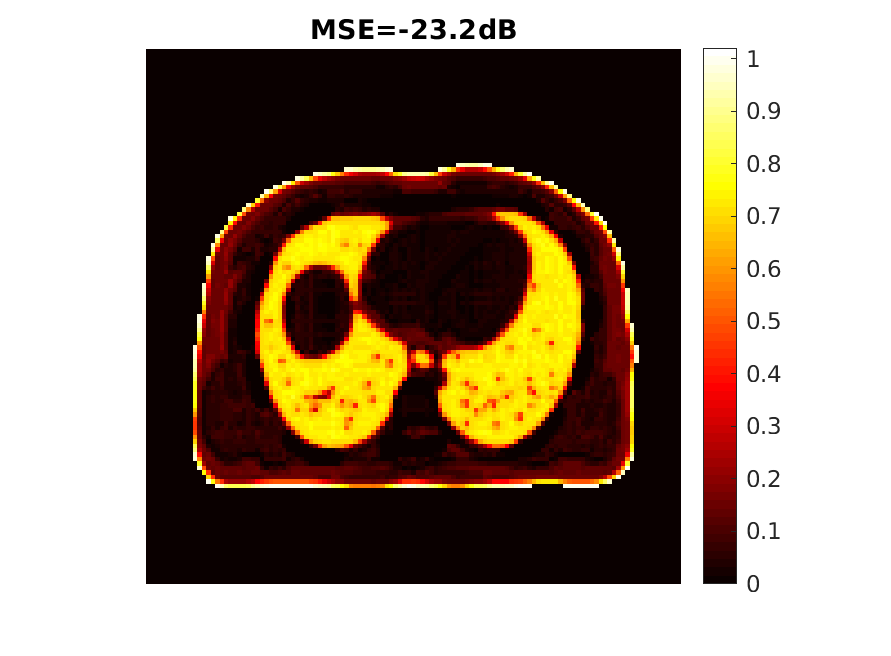}} \\
{\includegraphics[trim=2cm 1cm 3.5cm 0cm, clip, height=4.2cm]{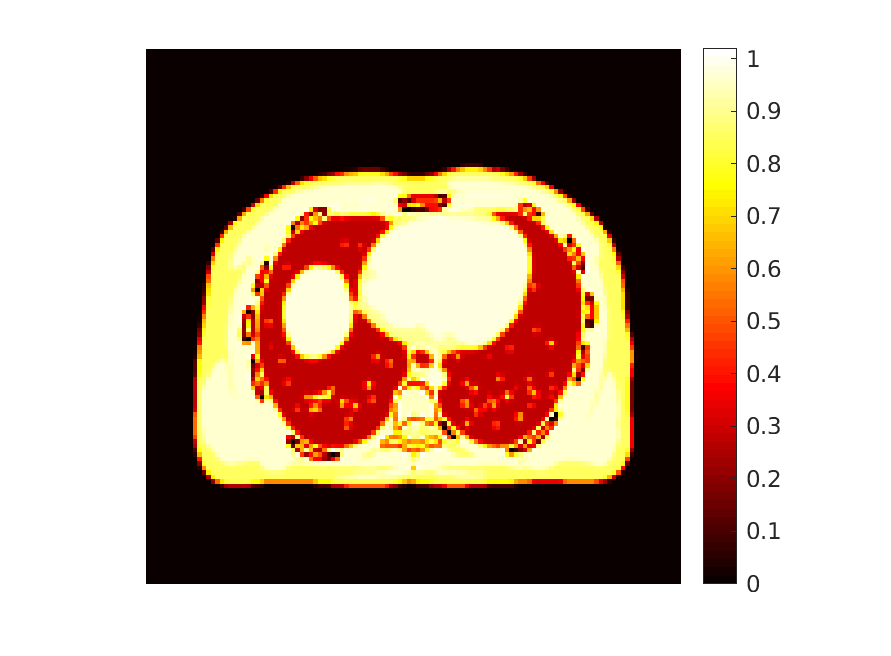}} &
{\includegraphics[trim=2cm 1cm 3.5cm 0cm, clip, height=4.2cm]{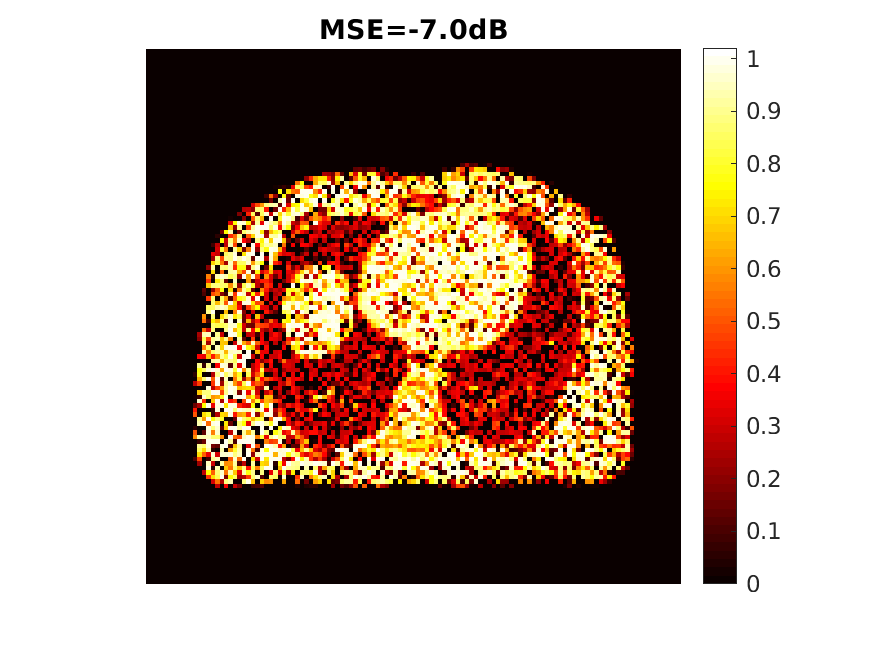}} &
{\includegraphics[trim=2cm 1cm 1.8cm 0cm, clip, height=4.2cm]{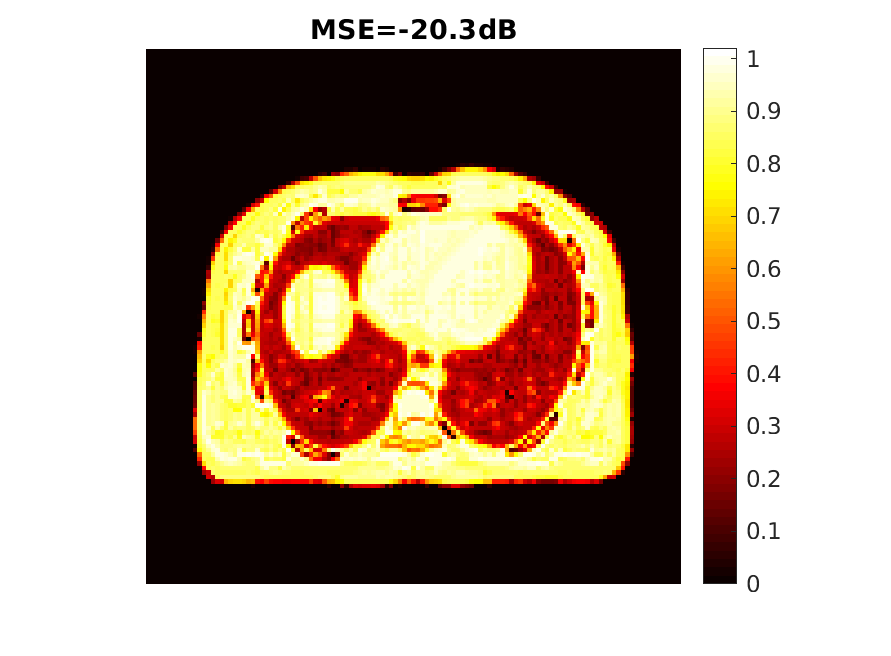}} \\
{\includegraphics[trim=2cm 1cm 3.5cm 0cm, clip, height=4.2cm]{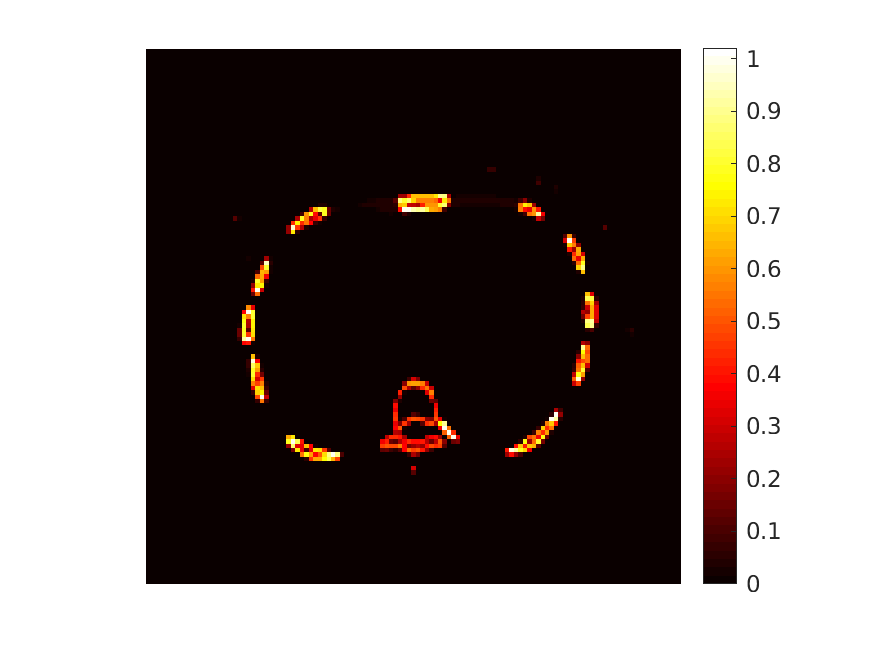}} &
{\includegraphics[trim=2cm 1cm 3.5cm 0cm, clip, height=4.2cm]{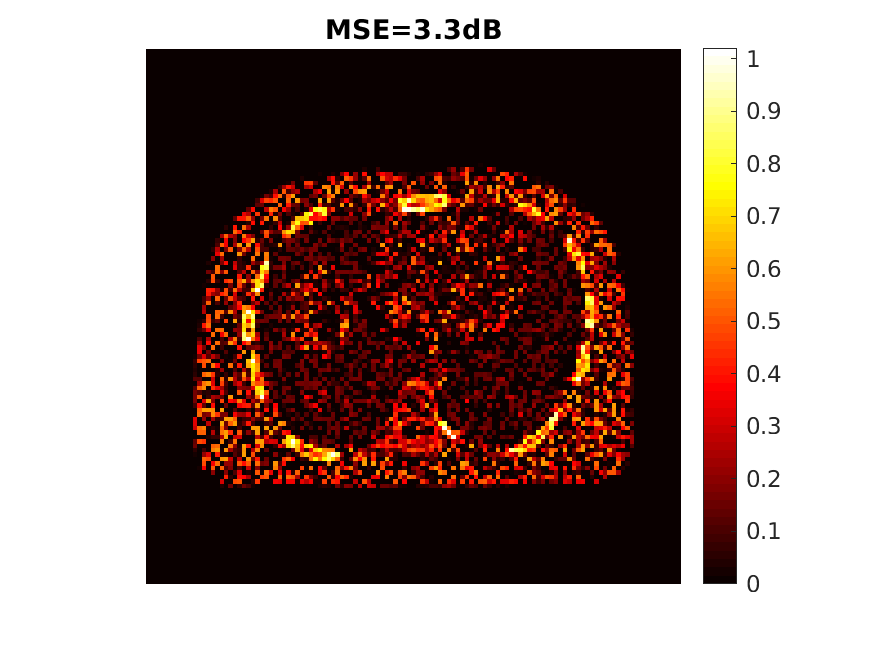}} &
{\includegraphics[trim=2cm 1cm 1.8cm 0cm, clip, height=4.2cm]{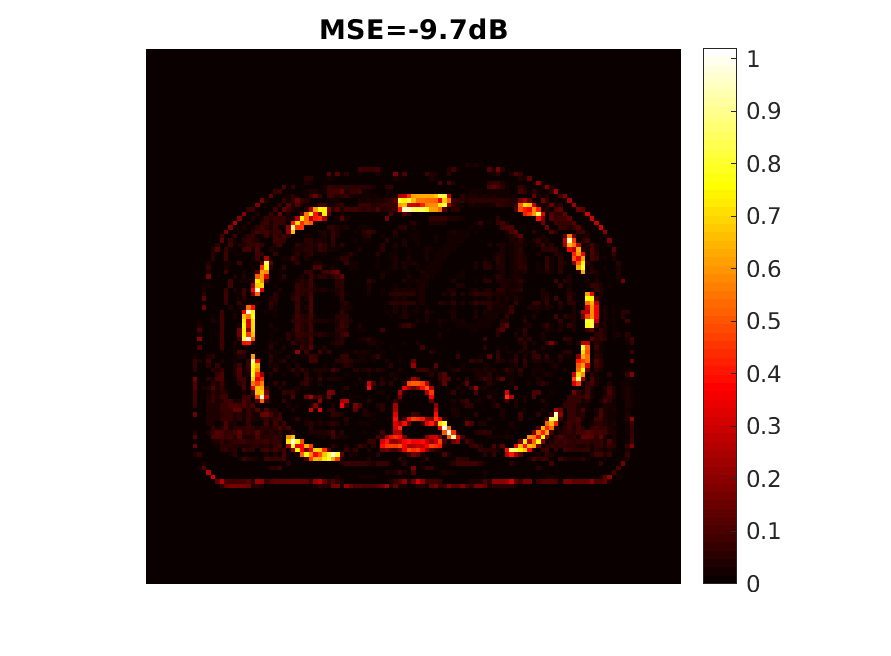}} \\
(a) Truth & (b) MLAA & (c) Kernel MLAA\\
\end{tabular}
\caption{True and estimated fractional images of three basis materials - air (top row), soft tissue (middle row), and bone (bottom row) - by different reconstruction algorithms.  (a) Ground truth; (b) Standard MLAA; (c) Proposed kernel MLAA. }
\label{fig-mmd-images}
\ec
\end{figure*}

\subsection{Comparison Results for GCT Image Quality} 

Fig. \ref{fig-rec-images} shows the reconstructed GCT attenuation images at 511 keV from the noisy PET emission data using the standard MLAA and proposed kernel MLAA algorithms with 400 iterations. Both the results of using the uniform initial and CT initial are shown.  It is not surprising that the CT initial provided better image quality because the initial estimate is closer to the ground truth. For both initials, the kernel MLAA achieved much better results with lower MSE than the standard MLAA reconstruction. 

Fig. \ref{fig-rec-mse} shows the resulting MSE as a function of iteration number in different reconstruction algorithms. The post-reconstruction denoising with  kernel smoothing (KS) is also included in the comparison. For all the three reconstruction algorithms, the image initials made a large difference at early iterations but not at late iterations where the image reconstructions start to converge despite the initial starting point. In all the three reconstruction approaches, the CT initial also allowed an earlier iteration stopping to get each own best MSE than the uniform initial. This is useful as less number of iteration leads to accelerated speed. While post-reconstruction denoising improved the MLAA result, the kernel MLAA achieved a larger improvement on image quality with lower MSE. Because the CT initial demonstrated better performance than the uniform initial, hereafter we mainly present further comparisons based on the CT initial. 

The effect of count level is shown in Fig. \ref{fig-rec-mse-count}.  In addition to the 5 million count level, two additional count levels (1 million and 10 million) were also included in the study. The number of iterations was fixed at 400 for each reconstruction. With increased count level, image quality by different algorithms were all increased. The kernel MLAA remains superior over the MLAA (with or without post-reconstruction smoothing) at different count levels.

Fig. \ref{fig-rec-BiasSD} shows the results of ensemble bias versus SD for GCT ROI quantification in a liver region and a spine bone region. The count level was 5 million events. The iteration number varies from 0 to 3000 with a step of 100 iterations. As iteration number increases, the bias of ROI quantification is reduced while the SD is increased. After a certain number of iterations, the increasing noise may become dominant, which in turn induces higher bias. The post-reconstruction kernel smoothing approach outperformed the standard MLAA approach  in a homogeneous region such as the liver but may oversmooth small targets such as the bone structures. The kernel MLAA achieved the best performance for both ROIs. At a fixed bias level, the kernel MLAA has lower SD than the other two approaches.

\subsection{Comparison Results for Multi-material Decomposition}

The results of applying multi-material decomposition (MMD) to the combined X-ray CT and GCT data are given in Figure \ref{fig-mmd-images}.  The MLAA and kernel MLAA reconstructions were run for 400 iterations. The ground truth of the three basis fractional images (air, soft tissue, bone) was generated using the noise-free pair of low-energy x-ray CT image and the 511 keV GCT image. The images by MLAA contain substantial noise but the regions of air and bone were still differentiated from the soft-tissue basis. Compared to MLAA, the kernel MLAA reconstruction led to a dramatic noise reduction in all the three basis images with increased image MSE. 

The MSE of each basis fractional image is further plotted as a function of iteration number in Fig. \ref{fig-mmd-images}(d) in which the post-reconstruction kernel smoothing approach was also included for comparison. The kernel MLAA approach demonstrated a significant MSE improvement over the conventional MLAA approach with or without post-reconstruction smoothing across all iterations. 

To demonstrate the performance of different reconstruction algorithms for ROI quantification on MMD images, Fig. \ref{fig-mmd-BiasSD} shows the bias versus SD trade-off plot for ROI quantification on the bone fractional image using the spine ROI as shown in Fig. \ref{fig-rec-BiasSD}(a). Due to over-smoothing, the post-reconstruction denoising approach had lower SD but higher bias, resulting in an even worse trade-off than the MLAA without denoising. In comparison, the kernel MLAA reconstruction achieved a consistently better trade-off than the other two approaches.

\begin{figure*}[t]
\bc\footnotesize
\begin{tabular}{c c c}
{\includegraphics[trim=0.5cm 0cm 1.3cm 0.5cm, clip, height=3.9cm]{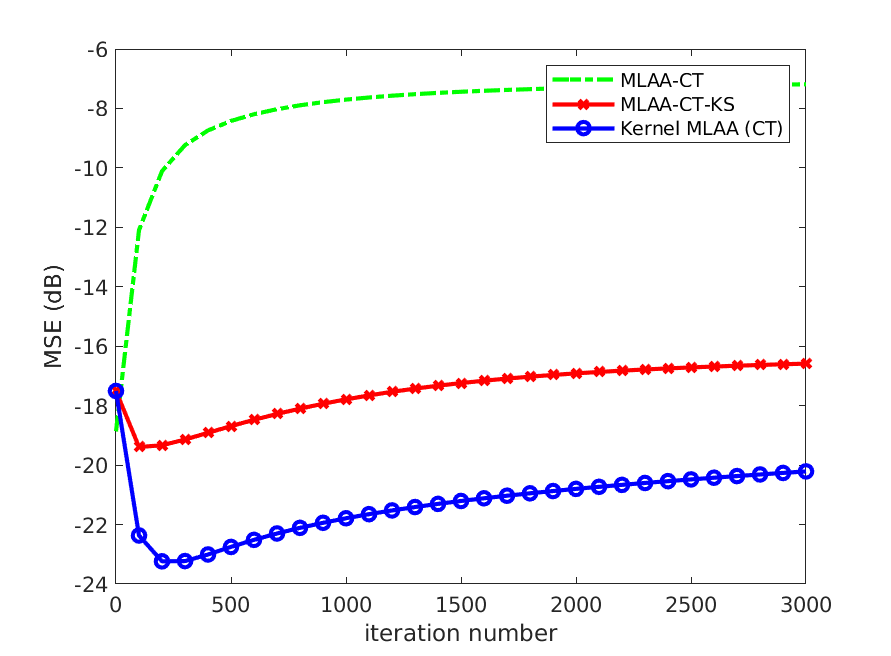}} &
{\includegraphics[trim=0.5cm 0cm 1.3cm 0.5cm, clip, height=3.9cm]{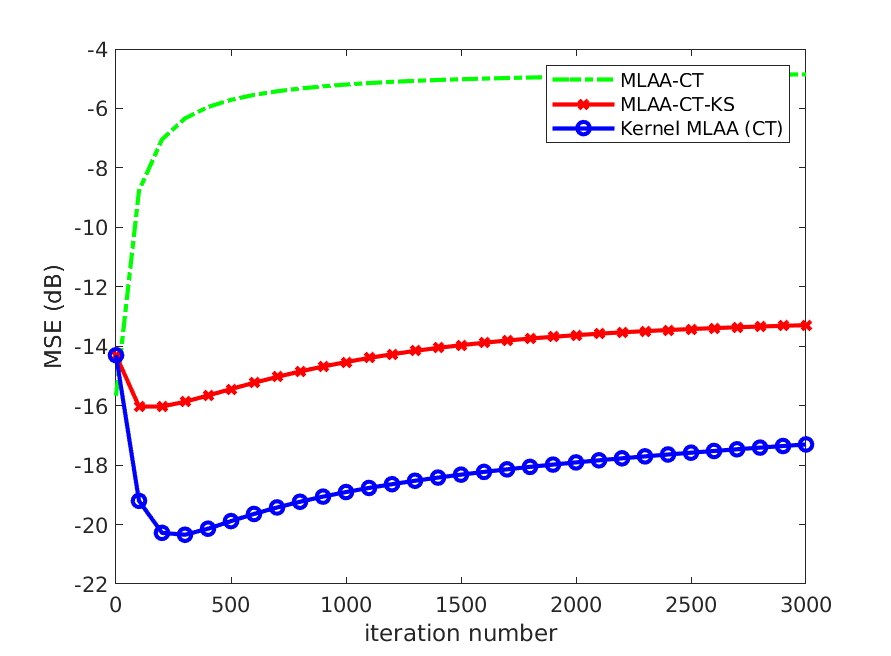}} &
{\includegraphics[trim=0.5cm 0cm 1.3cm 0.5cm, clip, height=3.9cm]{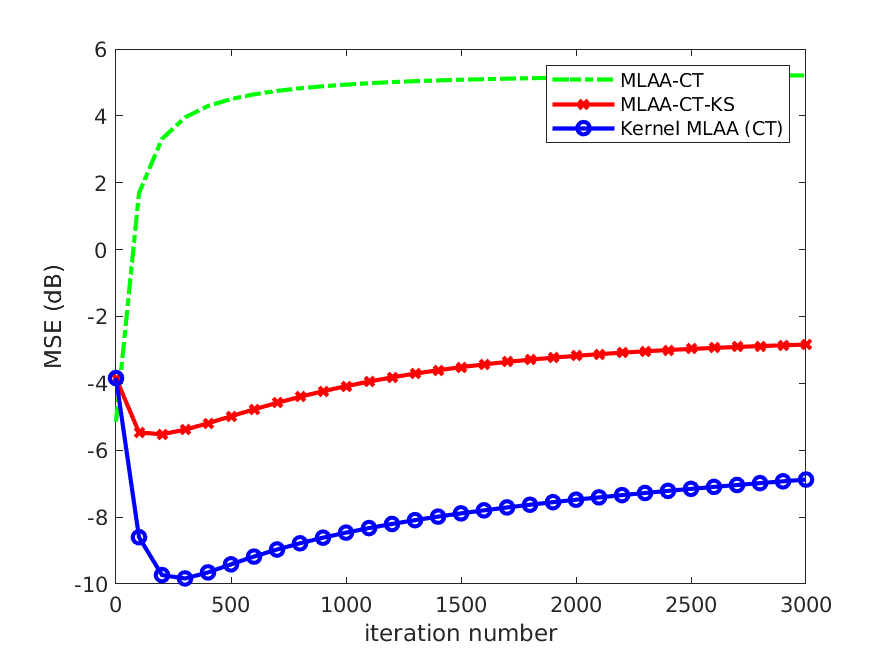}}\\
(a) air & (b) soft tissue & (c) bone\\ 
\end{tabular}
\caption{Plot of image MSE as a function of iteration number for each basis fractional image. }
\label{fig-mmd-mse}
\ec
\end{figure*}
 
\begin{figure}[t]
\bc\footnotesize
{\includegraphics[trim=0cm 0cm 1cm 0cm, clip, height=6.0cm]{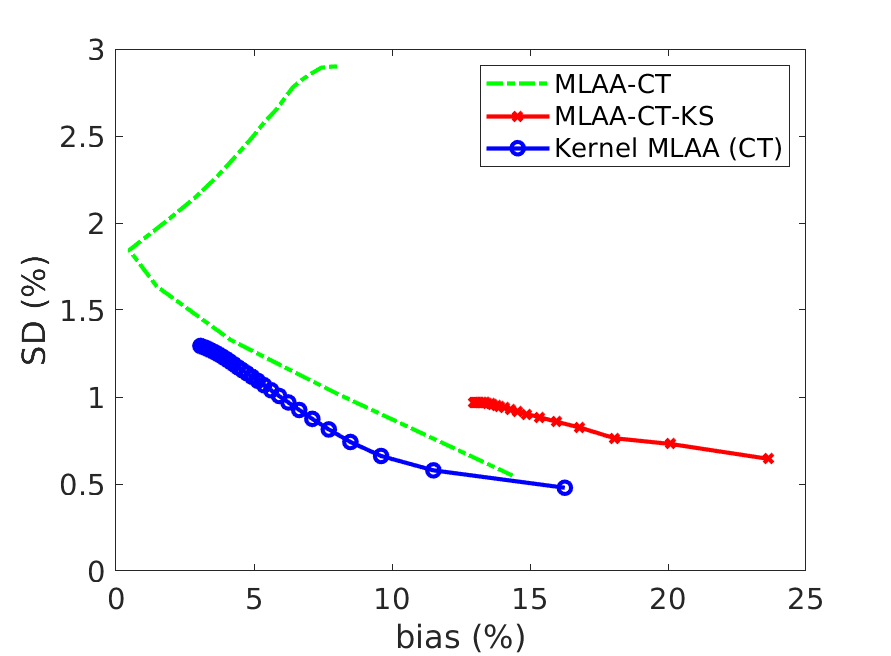}}
\caption{Plot of bias versus standard deviation trade-off for ROI quantification on the fractional image of bone basis material.}
\label{fig-mmd-BiasSD}
\ec
\end{figure}

\section{Discussions}

In this paper, we demonstrated the feasibility of PET-enabled dual-energy CT imaging using computer simulation. While the standard MLAA reconstruction suffers from high noise, the kernel MLAA reconstruction can dramatically improve the GCT image quality and multi-material decomposition by utilizing the X-ray CT image prior for MLAA reconstruction. This shows a promising direction and also provides guidance to test the method in future physical phantom experiments and real patient data studies that will be conducted as our next steps. 

In the current study, we consider the X-ray CT is perfect, which however is less the case in practice. Depending on the application scenarios, X-ray CT of PET/CT may be of high noise and artifacts. Solutions include improved X-ray CT image reconstruction or deep-learning low-dose CT denoising (e.g., \cite{Li20} and references therein). One of our future work will investigate the effect of X-ray CT image quality on kernel MLAA and material decomposition and how improved methods may conquer the problems. It is also possible to combine X-ray CT projection data and PET emission data to pursue joint MLAA/CT reconstruction from simultaneous emission-transmission scans, in a way similar to (while still different from) earlier and new effort on a related problem (e.g., \cite{Erdogan1999, Cheng20}). 

Another challenge down the road is that it is not uncommon that misalignment exists between a PET scan and an X-ray CT scan due to patient movement and physiological motion. This problem may affect both the kernel MLAA reconstruction and the match between GCT and X-ray CT for dual-energy imaging.  One solution is to register the X-ray CT image to the GCT image estimated by standard MLAA with post-smoothing. The kernel MLAA and material decomposition are then implemented based on the registered X-ray CT image.

Despite the challenges, the PET-enabled dual-energy CT method has many potentials. (1) It may allow dual-energy CT imaging on PET/CT with a lower radiation exposure due to one less X-ray CT scan. In the context of whole-body $^{18}$F-fluodeoxyglucose (FDG) PET/CT, the effective dose of a low-dose CT scan for attenuation correction and anatomic localization is about 3-10 mSv while the effective dose of 10 mCi FDG is 7 mSv. A second X-ray CT scan for dual-energy imaging may add a significant amount of radiation exposure; (2) It may enable multi-energy spectral imaging in two different ways. The first way is to combine dual-energy X-ray CT with the PET-enabled GCT at 511 keV to create triple-energy CT imaging. The second way is to derive another high-energy  GCT images from a PET scan in addition to the 511 keV attenuation image. This is possible because $^{176}$Lu in the LSO/LYSO crystals of PET detectors produces background radiation at 307 keV and 202 keV, which can be used to obtain the corresponding attenuation maps as demonstrated by Rothfuss {\em et. al.} \cite{Rothfuss14}; (3) In addition, the proposed PET-enabled dual-energy imaging method also has the potential to correct X-ray CT artifacts. X-ray CT is commonly poly-energetic and suffers from scattering \cite{Endo06} and beam hardening effects especially in the presence of metallic implants \cite{Gompel11}. The essentially mono-energetic 511 keV GCT enabled by PET could potentially help reduce the beam hardening, scattering, and metal artifacts of X-ray CT. We will explore these directions in our future work.
 
\section{Conclusion}  

We have developed a PET-enabled dual-energy CT imaging method and demonstrated its proof of concept using computer simulation. Distinct from conventional dual-energy CT imaging, the proposed method does not use two X-ray CT energies but combines low-energy X-ray CT and high-energy gamma-ray CT reconstructed from time-of-flight PET emission data. A kernel MLAA algorithm has also been developed to improve image quality and validated using simulated data. The results have shown the feasibility of the method for multi-material decomposition. As compared to a standard PET/CT scan, the proposed method can add a new dimension of information of material compositions without increasing the imaging time and cost. The method also has the potential to be extended for multi-energy spectral CT imaging.

 \section*{Acknowledgment}
   
This work is supported by National Institutes of Health (NIH) under grant no. R21 EB027346. The author thanks Dr. Jian Zhou for providing the time-of-flight forward and back projectors, Dr. Simon R. Cherry, Dr. Jinyi Qi, Dr. Ramsey D. Badawi, and Dr. Ben Spencer for helpful discussions.
 
\section*{References}
\bibliographystyle{apalike}

 \bibliographystyle{plain}

\begin{thebibliography}{1} 

\bibitem{McCollough15}	
McCollough CH, Leng SA, Yu LF, {\em et al.},
\newblock ``Dual- and Multi-Energy CT: Principles, Technical Approaches, and Clinical Applications,''
\newblock {\em Radiology}, 276:637-653, 2015. 

\bibitem{McCollough20}	
CH McCollough, K Boedeker,  D. Cody, {\em et al.},
\newblock ``Principles and applications of multienergy CT: Report of AAPM Task Group 291,''
\newblock {\em Medical Physics}, 247(7): e881-e912, 2020.

\bibitem{Noh09}	
Noh J, Fessler JA, Kinahan PE, 
\newblock ``Statistical sinogram restoration in dual-energy CT for PET attenuation correction,''
\newblock {\em EEE Transactions on Medical Imaging}, 28(11):1688-1702,2009.

\bibitem{Xia14}	
T. Xia, A. M. Alessio, P. E. Kinahan,
\newblock ``Dual energy CT for attenuation correction with PET/CT,''
\newblock {\em Medical Physics}, 41(1):012501, ,2014.

\bibitem{Cecco18}	
C. N. De Cecco, P. Burchett, M. van Assen, {\em et al.},
\newblock ``Rationale and design of a prospective study on the first integrated PET/dual-energy CT system for staging and image-based radiation therapy planning of lung cancer,''
\newblock {\em European Radiology Experimental}, 2:15, 2018.

\bibitem{Wu20}	
H. Wu, S. Dong, ,X. Li, {\em et al.},
\newblock ``Clinical utility of dual-energy CT used as an add-on to 18F FDG PET/CT in the preoperative staging of resectable NSCLC with suspected single osteolytic metastases,''
\newblock {\em Lung Cancer}, 140: 80-86, 2020.

\bibitem{Townsend08}	
Townsend DW, 
\newblock ``Multimodality imaging of structure and function,''
\newblock {\em Physics in Medicine and Biology}, 2008;53:R1-R39.

\bibitem{Defrise12}	
Defrise M, Rezaei A, Nuyts J,
\newblock ``Time-of-flight PET data determine the attenuation sinogram up to a constant,''
\newblock {\em Physics in Medicine and Biology}, 57:885-899 , 2012. 

\bibitem{Rezaei12}	
Rezaei A, Defrise M, Bal G, {\em et al.},
\newblock ``Simultaneous reconstruction of activity and attenuation in time-of-flight PET,''
\newblock {\em IEEE Transations on Medical Imaging}, 31:2224-2233, 2012. 


\bibitem{Nuyts18}	
J Nuyts,  A Rezaei, M Defrise,
\newblock ``The Validation Problem of Joint Emission/Transmission Reconstruction From TOF-PET Projections,''
\newblock {\em IEEE Transactions on Radiation and Plasma Medical Sciences }, 2(4):273- 278, 2018. 


\bibitem{Rezaei14}	
A. Rezaei, M. Defrise,  J. Nuyts
\newblock ``ML-Reconstruction for TOF-PET With Simultaneous Estimation of the Attenuation Factors,''
\newblock {\em IEEE Transations on Medical Imaging}, 33(7): 1563-1572, 2014. 

\bibitem{Defrise14}	
Defrise M, Rezaei A, Nuyts J,
\newblock ``Transmission-less attenuation correction in time-of-flight PET: analysis
of a discrete iterative algorithm,''
\newblock {\em Physics in Medine and Biology}, 59:1073-1095, 2014. 

\bibitem{Berker16}	
Berker Y, Li YS,
\newblock ``Attenuation correction in emission tomography using the emission data — A review,''
\newblock {\em Medical Physics}, 43:807-832, 2016. 

\bibitem{Li15}	
Li YS, Defrise M, Metzler SD, Matej S.,
\newblock ``Transmission-less attenuation estimation from time-of-flight PET histo-images using consistency equations,''
\newblock {\em Physics in Medicine and Biology}, 60:6563-6583, 2015.

\bibitem{Feng18}	
T Feng, J Wang, H Li,
\newblock ``Joint activity and attenuation estimation for PET with TOF data and single events,''
\newblock {\em Physics in Medicine and Biology}, 63(24): 245017, 2018.

\bibitem{Cheng20}
L Cheng,  T Ma,  X Zhang,  Q Peng,  Y Liu,  J Qi,
\newblock ``Maximum likelihood activity and attenuation estimation using both emission
and transmission data with application to utilization of Lu-176 background
radiation in TOF PET,''
\newblock {\em Medical Physics}, vol. 47, no. 3, pp.  1067-1082, 2020.

\bibitem{Panin13} 
V Y Panin, M Aykac and M E Casey,
\newblock ``Simultaneous reconstruction of emission activity and attenuation coefficient distribution from TOF data, acquired with external transmission source,'' 
\newblock {\em Phys. Med. Biol.}, 58: 3649, 2013.

\bibitem{Bousse16}	
A. Bousse, O. Bertolli, D. Atkinson, {\em et al.},
\newblock ``Maximum-Likelihood Joint Image Reconstruction/Motion Estimation in Attenuation-Corrected Respiratory Gated PET/CT Using a Single Attenuation Map,''
\newblock {\em IEEE Transactions on Medical Imaging}, 35(1): 217-228, 2016.

\bibitem{Presotto15}	
L. Presotto, E. Busnardo, D. Perani, {\em et al.},
\newblock ``Simultaneous reconstruction of attenuation and activity in cardiac PET can remove CT misalignment artifacts,''
\newblock {\em Journal of Nuclear Cardiology}, 23: 1086–1097, 2016.

\bibitem{Rezaei18} 
A Rezaei, C M. Deroose, T Vahle, F Boada, and J Nuyts,
\newblock ``Joint Reconstruction of Activity and Attenuation in Time-of-Flight PET: A Quantitative Analysis,'' 
\newblock {\em Journal of Nuclear Medicine}, 59(10): 1630-1635, 2018.

\bibitem{Mehranian15}	
A. Mehranian, H. Zaidi,
\newblock ``Joint Estimation of Activity and Attenuation in Whole-Body TOF PET/MRI Using Constrained Gaussian Mixture Models,''
\newblock {\em IEEE Transactions on Medical Imaging}, 34(9): 1808-1821, 2015.

\bibitem{Benoit16}	
D Benoit, C N Ladefoged, A Rezaei, {\em et al.},
\newblock ``Optimized MLAA for quantitative non-TOF PET/MR of the brain,''
\newblock {\em Physics in Medine and Biology}, 61: 8854, 2016. 

\bibitem{Heuber17}	
T Heußer, C M. Rank, YBerker,  {\em et al.},
\newblock ``MLAA-based attenuation correction of flexible hardware components in hybrid PET/MR imaging,''
\newblock {\em Physics in Medicine and Biology}, 4, 12, 2017.

\bibitem{Ahn18}	
S Ahn, L Cheng, D D Shanbhag,  {\em et al.},
\newblock ``Joint estimation of activity and attenuation for PET using pragmatic MR-based prior: application to clinical TOF PET/MR whole-body data for FDG and non-FDG tracers
,''
\newblock {\em Physics in Medicine and Biology}, 63(4): 045006, 2018.

\bibitem{Hwang18} 
Hwang D, Kim KY, Kang SK, Seo S, Paeng JC, Lee DS, Lee JS,
\newblock ``Improving the Accuracy of Simultaneously Reconstructed Activity and Attenuation Maps Using Deep Learning,'' 
\newblock {\em Journal of Nuclear Medicine}, 59(10): 1624 - 1629, 2018.

\bibitem{Rezaei19} 
A Rezaei, G Schramm, S. M.A. Willekens, G. Delso, K. Van Laere, and J. Nuyts,
\newblock ``A Quantitative Evaluation of Joint Activity and Attenuation Reconstruction in TOF PET/MR Brain Imaging,'' 
\newblock {\em Journal of Nuclear Medicine}, 60(11): 1649-1655, 2019.

\bibitem{Bowsher96}
J. E. Bowsher, V. E. Johnson, T. G. Turkington, R. J. Jaszczak, C. E. Floyd, R. E. Coleman,
\newblock ``Bayesian reconstruction and use of anatomical a priori information for emission tomography,''
\newblock {\em IEEE Transactions on Medical Imaging}, vol. 15, no. 5, pp. 673-686, 1996.

\bibitem{Wang2015} 
Wang G., Qi J., 
\newblock ``PET image reconstruction using kernel method,'' 
\newblock {\em IEEE Transactions on Medical Imaging}, vol. 34, no. 1, pp. 61-71, 2015.

\bibitem{Hutchcroft16} 
Hutchcroft W., Wang G., Chen K., Catana C., Qi J., 
\newblock ``Anatomically-aided PET reconstruction using the kernel method,'' 
\newblock {\em Physics in Medicine and Biology}, 61(18): 6668-6683, 2016.

\bibitem{Bland18} 
J. Bland, A. Mehranian, M. A. Belzunce, S. Ellis, C. J. McGinnity, A. Hammers, A. J. Reader,
\newblock ``MR-guided kernel EM reconstruction for reduced dose PET imaging,'' 
\newblock {\em IEEE Transactions on Radiation and Plasma Medical Sciences }, 2(3): 235 - 243, 2018.

\bibitem{Novosad16} 
P. Novosad, A. J. Reader, 
\newblock ``MR-guided dynamic PET reconstruction with the kernel method and spectral temporal basis functions,'' 
\newblock {\em Physics in Medicine and Biology}, 61(12): 4624–4645, 2016.

\bibitem{Gong18} 
K. Gong, J. Cheng-Liao, G. B. Wang, K. T. Chen, C. Catana, J. Qi,
\newblock ``Direct Patlak reconstruction from dynamic PET data using kernel method with MRI information based on structural similarity,'' 
\newblock {\em IEEE Transactions on Medical Imaging}, 37(4): 955-965, 2018.

\bibitem{Deidda19} 
D Deidda, NA Karakatsanis, PM Robson,  {\em et al.},
\newblock ``Hybrid PET-MR list-mode kernelized expectation maximization reconstruction,'' 
\newblock {\em Inverse Problems}, 35(4):  044001, 2019.

\bibitem{Wang2019} 
Wang G., 
\newblock ``High temporal-resolution dynamic PET image reconstruction using a new spatiotemporal kernel method,'' 
\newblock {\em IEEE Transactions on Medical Imaging}, 38(3): 664 – 674, 2019.

\bibitem{Nuyts1999} 
J. Nuyts, P. Dupont, S. Stroobants, {\em et al.},
\newblock ``Simultaneous maximum a posteriori reconstruction of attenuation and activity distributions from emission sinograms,'' 
\newblock {\em IEEE Transactions on Medical Imaging}, 18(5): 393 - 403, 1999.

\bibitem{Wang18} 
Wang GB,
\newblock ``PET-enabled dual-energy CT: A proof-of-concept simulation study,'' 
\newblock {\em Conference Proceedings of 2018 IEEE Nuclear Science Symposium and Medical Imaging  (NSS\&MIC)}, Sydney, Australia, November 13-17, 2018. DOI: 10.1109/NSSMIC.2018.8824351.

\bibitem{Qi2006} 
J. Qi, R. M. Leahy,
\newblock ``Iterative reconstruction techniques in emission computed tomography,'' 
\newblock {\em Physics in Medicine and Biology}, 51:R541-578, 2006.

\bibitem{Kinahan03}
Kinahan PE, Hasegawa BH, Beyer T
\newblock ``X-ray-based attenuation correction for positron emission tomography/computed tomography scanners,''
\newblock {\em Seminars in Nuclear Medicine}, 33(3): 166–179, 2003. 

\bibitem{Dicken1999} 
V Dicken,
\newblock ``A new approach towards simultaneous activity and attenuation reconstruction in emission tomography,'' 
\newblock {\em Inverse Problems}, 15(4):  931-960, 1999.

\bibitem{Salomon2011} 
A Salomon, A Goedicke, B Schweizer, T Aach, V Schulz
\newblock ``Simultaneous Reconstruction of Activity and Attenuation for PET/MR,'' 
\newblock {\em IEEE Transactions on Medical Imaging}, 30(3):  804 - 813, 2011.

\bibitem{Conti2011} 
M. Conti,
\newblock ``Why is TOF PET reconstruction a more robust method in the presence of inconsistent data?,'' 
\newblock {\em Physics in Medicine and Biology}, 56(1): 155–168, 2011.

\bibitem{Erdogan99}
H Erdogan, JA Fessler,
\newblock ``Monotonic algorithms for transmission tomography,''
\newblock {\em IEEE Transactions on Medical Imaging}, vol. 18, no. 9, pp. 801--814, April 1999. 

\bibitem{Erdogan99b}
H Erdogan, JA Fessler,
\newblock ``Ordered subsets algorithms for transmission tomography,''
\newblock {\em Physics in Medicine and Biology}, vol. 44, no. 11, pp. 2835-51, 1999.


\bibitem{Li20}
Li SQ, Wang GB,
\newblock ``Low-dose CT image denoising using parallel-clone networks,''
\newblock {\em arXiv}, eprint 2005.06724, pp. 1-10, May 2020.


\bibitem{Erdogan1999} 
H. Erdogan, J.A. Fessler,
\newblock ``Joint estimation of attenuation and emission images from PET scans,'' 
\newblock {\em Conference Proceedings of 1999 IEEE Nuclear Science Symposium and Medical Imaging  (NSS\&MIC)}, Seattle WA, 24-30 Oct., 1999. DOI: 10.1109/NSSMIC.1999.842904.


\bibitem{Rothfuss14} 
H Rothfuss, V Panin, A Moor, J Young, {\em et al.},
\newblock ``LSO background radiation as a transmission source using time of flight,'' 
\newblock {\em Physics in Medicine and Biology}, 59(18): 5483-500, 2014.

\bibitem{Endo06} 
Endo M, Mori S, Tsunoo T, Miyazaki H., 
\newblock ``Magnitude and effects of x-ray scatter in a 256-slice CT scanner,'' 
\newblock {\em Medical Physics}, 59(18): 33:3359-3368, 2016.

\bibitem{Gompel11} 
Van Gompel G, Van Slambrouck K, Defrise M, {\em et al.},
\newblock ``Iterative correction of beam hardening artifacts in CT,'' 
\newblock {\em Medical Physics}, 38:S36-S49, 2011.

\end{thebibliography}

\end{document}